\documentclass[twoside,11pt]{article}
\usepackage{jmlr2e}
\usepackage{centernot}
\usepackage{epsfig}
\usepackage{graphicx}
\usepackage{amssymb,amsmath}
\usepackage{verbatim,enumerate}
\usepackage{verbatim,multicol,color}
\usepackage{multicol}
\usepackage[english]{babel}
\usepackage[position=top]{subfig}
\usepackage{color}
\usepackage{float}
\usepackage{hyperref}
\hypersetup{
     colorlinks=true,
     linkcolor=blue,
     citecolor=blue,
     filecolor=blue,
     urlcolor=blue,
}
\usepackage{setspace}
\usepackage{mathrsfs}
\usepackage[skip=0pt]{caption}
\usepackage{rotating}
\usepackage{pdflscape}

\def\log{\hbox{log}}

\def\boxit#1{\vbox{\hrule\hbox{\vrule\kern6pt
          \vbox{\kern6pt#1\kern6pt}\kern6pt\vrule}\hrule}}

\def\bse{\begin{eqnarray*}}
\def\ese{\end{eqnarray*}}
\def\be{\begin{eqnarray}}
\def\ee{\end{eqnarray}}
\def\bq{\begin{equation}}
\def\eq{\end{equation}}
\def\bse{\begin{eqnarray*}}
\def\ese{\end{eqnarray*}}
\def\Pr{\hbox{P}}

\newcommand{\bI}{\mathbf{I}}

\newcommand{\bX}{\mathbf{X}}

\newcommand{\bZ}{\mathbf{Z}}

\newcommand{\bU}{\mathbf{U}}
\newcommand{\bV}{\mathbf{V}}

\newcommand{\bW}{\mathbf{W}}

\newcommand{\bH}{\mathbf{H}}
\newcommand{\bx}{\mathbf{x}}

\newcommand{\bu}{\mathbf{u}}

\newcommand{\by}{\mathbf{y}}

\newcommand{\bv}{\mathbf{v}}

\newcommand{\bmu}{\boldsymbol{\mu}}

\newcommand{\bbeta}{\boldsymbol{\beta}}

\newcommand{\0}{\mathbf{0}}
\newcommand{\1}{\mathbf{1}}
\newcommand*{\QEDB}{\hfill\ensuremath{\blacksquare}}
\newcommand{\rchi}{\mathscr X}
\newcommand{\argmin}{\operatornamewithlimits{arg\,min}}

\numberwithin{equation}{section}
\newtheorem{thm}{Theorem}[section]
\newtheorem{cor}[thm]{Corollary}
\newtheorem{lem}[thm]{Lemma}
\newtheorem{ex}{Example}

 \newtheorem{rmk}[thm]{Remark}

\newcommand{\J}{\mathbf{\J}}

\allowdisplaybreaks

\usepackage{lastpage}
\jmlrheading{22}{2021}{1-41}{10/20; Revised 5/21}{12/21}{20-1219}{Sarbojit Roy, Soham Sarkar, Subhajit Dutta and Anil K. Ghosh}

\ShortHeadings{Generalizations of distance based classifiers for HDLSS data}{Roy, Sarkar, Dutta and Ghosh}
\firstpageno{1}

\begin{document}

\title{On Generalizations of Some Distance Based Classifiers\\ for HDLSS Data}
\vspace{0.1in}

\author{\name Sarbojit Roy \email sarbojit@iitk.ac.in \\
	\addr Department of Mathematics and Statistics\\
	IIT Kanpur\\
	Kanpur - 208016, India.
	\AND
	\name Soham Sarkar \email soham.sarkar@epfl.ch \\
	\addr Institut de Math{\'e}matiques\\
	{\'E}cole Polytechnique F{\'e}d{\'e}rale de Lausanne\\
	1015 Lausanne, Switzerland.
	\AND
	\name Subhajit Dutta \email duttas@iitk.ac.in \\
	\addr Department of Mathematics and Statistics\\
	IIT Kanpur\\
	Kanpur - 208016, India.
	\AND
	\name Anil K. Ghosh \email akghosh@isical.ac.in \\
	\addr Theoretical Statistics and Mathematics Unit\\
	Indian Statistical Institute\\ 
	Kolkata - 700108, India.
	}	

\editor{Charles Elkan}

\maketitle

\begin{abstract}%
\vspace*{0.05in}

\noindent
In high dimension, low sample size (HDLSS) settings, classifiers based on Euclidean distances like the nearest neighbor classifier and the average distance classifier perform quite poorly if differences between locations of the underlying populations get masked by scale differences. To rectify this problem, several modifications of these classifiers have been proposed in the literature. However, existing methods are confined to location and scale differences only, and they often fail to discriminate among populations differing outside of the first two moments. In this article, we propose some simple transformations of these classifiers resulting in improved performance even when the underlying populations have the same location and scale. We further propose a generalization of these classifiers based on the idea of grouping of variables. 
High-dimensional behavior of the proposed classifiers is studied theoretically. Numerical experiments with a variety of simulated examples as well as an extensive analysis of benchmark data sets from three different databases exhibit advantages of the proposed methods.
\vspace*{0.1in}
\end{abstract}


\newpage
\section{Introduction} \label{Intro}

Classification is a common task in machine learning. Given $n$ data points in $\mathbb{R}^d$ belonging to $J (\geq 2)$ classes, the goal of a classifier is to assign a class label to a new data point. In particular, distance based classifiers have gained popularity because they are quite simple, and easy to implement. Well-known classifiers such as the nearest neighbor classifier, the centroid classifier, and the average distance classifier use only the distance between observations to classify a new test case \citep[see, e.g.,][]{hastie2009elements,CH2009}. These classifiers also have nice theoretical properties. Under appropriate conditions, misclassification probabilities of these classifiers converge to the Bayes risk (in other words, {\it Bayes risk consistency}) as the training sample size increases \citep[see, e.g.,][]{DGL2013}.

In today's world, high-dimensional problems are frequently encountered in scientific areas like microarray gene expression studies, medical image analysis, spectral measurements in chemometrics, etc. A distinct characteristic of some of these problems is the presence of a very large number of features (or, data dimension) with a much smaller sample size. 
In such high dimension, low sample size (HDLSS) situations, Euclidean distance based classifiers face some natural drawbacks due to \emph{distance concentration} \citep[see, e.g.,][]{aggarwal2001surprising,francois2007concentration}. 
In \citet{HMN05}, the authors studied the effect of distance concentration on some popular classifiers based on Euclidean distances such as the centroid classifier and the nearest neighbor classifier, and derived conditions under which these classifiers yield \emph{perfect classification} in the HDLSS setup. We now give some insight into the idea of distance concentration in HDLSS scenarios.

Consider a random sample $\rchi_j=\{\bX_{j1},\ldots,\bX_{jn_j}\}$ of size $n_j$ from the $j$-th population for $1 \leq j \leq J$. We assume that these $n_j(\ge 2)$ observations are independent and identically distributed (i.i.d.) from a distribution function $\mathbf{F}_j$ on $\mathbb{R}^d$.
Define $\rchi = \cup_{j=1}^J \rchi_j$ to be the full training sample of size $n=\sum_{j=1}^J n_j$. For simplicity of analysis, we take $J=2$. Let $\bmu_{jd}$ and $\Sigma_{jd}$ denote the $d$-dimensional location vector and the $d \times d$ scale matrix, respectively, corresponding to $\mathbf{F}_j$ for $j=1,2$. Also, assume that the following limits exist: 
\begin{align*}
&\nu_{12}^2:=\lim_{d \to \infty} \big\{d^{-1}\|{\bmu}_{1d}-{\bmu}_{2d}\|^2\big\} \text{ and } \sigma_j^2=\lim_{d \to \infty} \big\{d^{-1}{\rm tr}(\Sigma_{jd})\big\} \mbox{ for } j=1,2.
\end{align*}
Here, $\|\cdot\|$ denotes the Euclidean norm on $\mathbb{R}^d$ and ${\rm tr}(A)$ is the sum of the diagonal elements of a $d \times d$ matrix $A$. The constants $\nu_{12}^2$ and $\sigma_1^2,\sigma_2^2$ are measures of the location difference and scales, respectively. 
In \citet{HMN05}, the authors showed that in the HDLSS asymptotic regime (when $n$ is fixed and $d$ goes to infinity), if $\nu_{12}^2<|\sigma_1^2-\sigma_2^2|$, the nearest neighbor (NN) classifier \emph{assigns all observations to the population having a smaller dispersion}. Later, \citet{CH2009} showed that the average distance (AVG) classifier is also \emph{useless} in such a scenario. 
In other words, Euclidean distance based classifiers may not yield satisfactory performance for high-dimensional data if the location difference is masked by the scale difference. To address this specific problem, some modifications of these classifiers have been proposed in the literature. \citet{CH2009} identified $|\sigma^2_{1}-\sigma^2_{2}|$ as a nuisance parameter, and proposed a scale adjustment to the discriminant of the average distance classifier. A non-linear transformation of the covariate space followed by NN classification was proposed by \citet*{DG2016}, while \citet{PMG2016} developed a NN classifier based on a new dissimilarity index. However, all these modified classifiers are known to perform well in the HDLSS setup under conditions like `$\nu^2_{12}>0$' or `either $\nu^2_{12}>0$ or $\sigma^2_1 \neq \sigma^2_2$'. To summarize, all the existing classifiers are particularly useful in high-dimensional spaces when the underlying distributions differ either in their locations and/or scales. Our interest is to analyze the performance of these classifiers under more general scenarios (in particular, when $\nu_{12}^2=0$ and $\sigma_1^2=\sigma_2^2$). We demonstrate this by considering some classification problems involving two populations.

\begin{ex}\label{ex1}
We consider two populations where the $d$ component variables are i.i.d. For the first population, the component distribution is $N(0,5/3)$, while it is $t_5$ for the second population. Here, $N(\mu,\sigma^2)$ denotes the univariate Gaussian distribution with mean $\mu$ and variance $\sigma^2$, and $t_{\nu}$ denotes the standard Student's $t$ distribution with $\nu$ degrees of freedom.
\end{ex}

\begin{ex}\label{ex2}
The two populations under consideration have the $d$-dimensional Gaussian distributions $N_d(\0_d, \Sigma_{1d})$ and $N_d(\0_d, \Sigma_{2d})$, where $\0_d$ is the $d$-dimensional vector of zeros, and $\Sigma_{1d}$ and $\Sigma_{2d}$ are {\it block diagonal} dispersion matrices having the following form:
\begin{center}
	$\Sigma_{jd}$ =
	$\begin{bmatrix}
	\bH_j & 0 & \cdots & 0 \\
	0 & \bH_j & \cdots & 0 \\
	\vdots &&\ddots & \vdots\\
	0 & \cdots & 0 &\bH_j\\
	\end{bmatrix}
	$ with 
	$\bH_j$ =
	$\begin{bmatrix}
	1&\rho_j\cdots \rho_j\\
	\rho_j& 1\cdots\rho_j\\
	\vdots &\vdots \ddots \vdots\\
	\rho_j&\rho_j\cdots 1\\
	\end{bmatrix}
	$ for $j=1,2$.\\
	\vspace{0.25cm}
\end{center}
In this example, we keep the size of the blocks fixed at ten (i.e., $\bH_j$ is a $10 \times 10$ matrix for $j=1,2$) and choose $\rho_1=0.3$ and $\rho_2=0.7$. 
\end{ex}

\begin{ex}\label{ex3}
We consider $d$-dimensional Gaussian distributions $N_d(\0_d,\Sigma_{1d})$ and $N_d(\0_d,\Sigma_{2d})$, where $\Sigma_{1d}$ and $\Sigma_{2d}$ have an {\it auto-regressive} covariance structure (i.e., $\Sigma_d = ((\rho^{|i-j|}))_{1 \le i,j \le d}$ and $0<\rho<1$) with parameters $0.3$ and $0.7$, respectively.
\end{ex}

\begin{figure}[htp]
\begin{center}
\captionsetup{justification=centering}
\includegraphics[width=\linewidth]{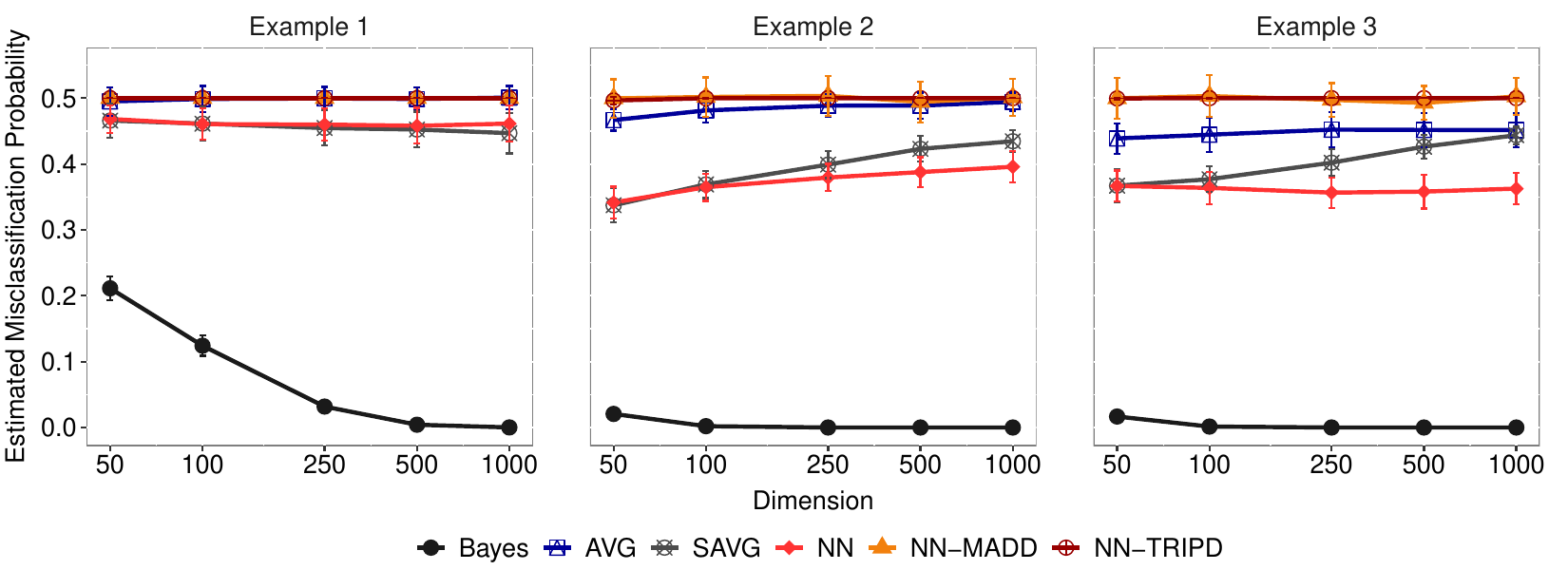}
\end{center}
\caption{Average misclassification rates (along with the standard errors) based on $100$ repetitions of various classifiers are plotted for increasing values of $d$ (in logarithmic scale). The classifiers AVG/SAVG, NN-MADD and NN-TRIPD were proposed by \cite{CH2009}, \cite{PMG2016} and \cite{DG2016}, respectively.} 
\label{plot0}
\end{figure}

For each example, we generated $50$ observations from each class to form the training sample. Misclassification rates of different classifiers are computed based on a test set consisting of $500$ ($250$ from each class) observations. This process was repeated $100$ times, and the average misclassification rates (along with the standard errors) of different classifiers for varying values of $d$ are shown in Figure \ref{plot0}. The Bayes risk was calculated for each example by computing the average Bayes risk over several random replicates of the data.
It is clear from Figure \ref{plot0} that none of the existing classifiers performed satisfactorily in these three examples. Observe that in all three examples, we have $\nu_{12}^2=0$ (the mean vectors $\bmu_{1d}$ and $\bmu_{2d}$ are equal to $\0_d$) and $\sigma_1^2=\sigma_2^2$ (both $\Sigma_{1d}$ and $\Sigma_{2d}$ have the same trace). This was the main reason behind the poor performance of all the existing classifiers.

In this article, we propose a modification to the Euclidean distance, and use it on two different distance based classifiers, namely, the scale-adjusted average distance classifier (henceforth referred to as SAVG) by \citet*{CH2009} and the NN classifier based on mean absolute differences of distances (henceforth referred to as NN-MADD) by \cite{PMG2016}. We show that these two classifiers, when used with the modified distance, can discriminate between populations even when there are no differences between their locations and scales. 
To capture discriminatory information, these modified distance based classifiers rely on the non-parametric concept of {\it energy} \citep[see][]{SR2017}.
In particular, if the one-dimensional marginals of the underlying populations are different, the proposed classifiers are shown to yield {\em perfect classification} in the HDLSS asymptotic regime. For HDLSS asymptotics, we fix the sample size $n$ and allow the data dimension $d$ to grow to infinity, which is different from standard asymptotics (with $d$ fixed and $n$ going to infinity).

The article is organized as follows. We define the modified classifiers and study their asymptotic properties in Section~\ref{general}.
In Section~\ref{Improve}, we propose further generalization of these classifiers for the case when the populations have same univariate marginals, but differ in their joint distributional structures (see Examples~\ref{ex2} and \ref{ex3}) and derive their asymptotic properties under the HDLSS setup. For implementation of the second generalization, we need to group the component variables into disjoint clusters.
In Section \ref{cluster}, we propose some data driven methods for this `variable clustering'. Numerical performance of the proposed classifiers on several simulated and real data sets are demonstrated in Sections \ref{sim} and \ref{real}, respectively. The article ends with a discussion in Section \ref{conclude}. 
All proofs and other mathematical details are provided in Appendix \ref{AppendixA}, and some additional material is presented as a \href{https://www.dropbox.com/sh/6wcdxzed8wg2xs3/AABag_Zp9biMOQQwt6XFaiF9a?dl=0}{Supplementary}.
A list of notations used in this paper is given in Appendix \ref{AppendixB}.

\section{Classifiers Based on Generalized Distances} \label{general}

Limitations of the classifiers discussed in the previous section stems from the fact that the behavior of the Euclidean distance in the HDLSS asymptotic regime is completely governed by the constants $\nu^2_{12}$, $\sigma_{1}^2$ and $\sigma_{2}^2$ \citep[see][]{HMN05}. As a consequence, Euclidean distance based classifiers cannot distinguish between populations that do not have differences in their first two moments. To circumvent this problem, we define a class of dissimilarity measures. For vectors $\bu=(u_1,\ldots,u_d)^{\top}$ and $\bv=(v_1,\ldots,v_d)^{\top}$, we define the dissmimilarity function $h_d^{\phi,\gamma}:\mathbb{R}^d \times \mathbb{R}^d \to \mathbb{R}^+$ between $\bu$ and $\bv$ as follows:
\vspace{-0.1cm}
\begin{equation}\label{modifyL2}
h_d^{\phi,\gamma}(\bu,\bv) \equiv h_d(\bu,\bv)=\phi \bigg(\frac 1 d \sum_{i=1}^d\gamma\big(|u_i - v_i|^2\big) \bigg),
\vspace{-0.1cm}
\end{equation}
where $\gamma:\mathbb{R}^+ \to \mathbb{R}^+$ and $\phi:\mathbb{R}^+ \to \mathbb{R}^+$ are continuous, monotonically increasing with $\gamma(0)=\phi(0)=0$. 
The class of functions \eqref{modifyL2} was proposed and used in the context of two-sample testing in \cite{sarkar2018}. It is interesting to note that if $\gamma (t)=t^{p/2}$ and $\phi(t)=t^{1/p}$ with $p>0$, then $h_d(\bu,\bv)$ is the $\ell_p$ distance (up to a constant involving $d$) between $\bu$ and $\bv$. This in particular includes the Euclidean distance (for $p=2$) as a special case.
In general, $h_d(\bu,\bv)$ need not be a distance function, but rather a measure of dissimilarity between $\bu$ and $\bv$. Our main objective is to use $h_d(\bu,\bv)$ instead of the {\em scaled} Euclidean distance (i.e., $d^{-1}\|\bu-\bv\|^2$ or $d^{-1/2}\|\bu-\bv\|$) in the SAVG and NN-MADD classifiers, and study their performance, both theoretically as well as numerically.

\subsection{Generalization of SAVG Classifier}

For a $J$-class problem and a new observation $\bZ$, the average distance (AVG) classifier is defined as 
\begin{align}
\delta_{\rm AVG}(\bZ)=\argmin_{1 \leq j \leq J} \bigg\{ \frac{1}{n_j} \sum_{\bX\in\rchi_j}d^{-1}\|\bX-\bZ\|^2 \bigg\}.
\end{align}

\noindent
If $\nu^2_{jj^\prime}>|\sigma^2_j-\sigma^2_{j^\prime}|$ for all $1\leq j\neq j^\prime\leq J$, then this classifier yields perfect classification in the HDLSS setup (i.e., the misclassification probability of the classifier goes to zero as $d\to\infty$, see \citealp{CH2009}). 
But, if this condition is violated, then this classifier may behave erratically by assigning all observations to the class having the smallest variance. To relax the condition stated above, the authors identified $|\sigma^2_j-\sigma^2_{j^\prime}|$ as a nuisance parameter, and proposed a scale adjustment to the average of distances as follows:
\begin{align}
&\xi^{(0)}_{jd}(\bZ)=\frac{1}{n_j} \sum_{\bX\in\rchi_j} d^{-1} \|\bX-\bZ\|^2-D^{(0)}_d(\rchi_j|\rchi_j)/2,
\end{align}
where $D^{(0)}_d(\rchi_j|\rchi_j)=\{n_j(n_j-1)\}^{-1} \sum_{\bX,\bX^{\prime}\in\rchi_j} d^{-1}\|\bX-\bX^{\prime}\|^2$ for all $1 \leq j \leq J$.
The scale-adjusted average distance (SAVG) classifier is defined as 
$$\delta_{\rm SAVG}(\bZ)=\argmin_{1 \leq j \leq J} \xi^{(0)}_{jd}(\bZ).$$ 
If $\nu^2_{jj^\prime}>0$ for all $1 \leq j\neq j^{\prime} \leq J$, then the misclassification probability of the SAVG classifier goes to zero as $d\to\infty$ \citep[see][Theorem~1]{CH2009}. The optimality condition for the SAVG classifier is clearly weaker than the one related to the AVG classifier. In other words, if the competing populations have difference only in their location parameters (irrespective of their differences in scales), the SAVG classifier perfectly classifies a new data point in high dimensions. However, we have observed deteriorating performance of the SAVG classifier in Figure \ref{plot0} when this condition is violated (recall that $\nu^2_{12}=0$ in Examples~\ref{ex1}, \ref{ex2} and \ref{ex3}). 

We modify the SAVG classifier by simply replacing the Euclidean distance $d^{-1}\|\bu-\bv\|^2$ with the new dissimilarity index $h_d(\bu,\bv)$, as stated below:
\begin{align}\label{SgAVG}
&\xi^{\phi,\gamma}_{jd}(\bZ) \equiv \xi_{jd}(\bZ)=\frac{1}{n_j} \sum_{\bX\in\rchi_j}h_d(\bZ,\bX)-D_d(\rchi_{j}|\rchi_j)/2.
\end{align}
Here, $D_d(\rchi_{j}|\rchi_j)=\{n_j(n_j-1)\}^{-1} \sum_{\bX,\bX^{\prime}\in\rchi_j} h_d(\bX,\bX^{\prime})$ with $n_j\geq 2$ for $1 \leq j \leq J$. The generalized scale-adjusted average distance (gSAVG) classifier based on $\xi_{jd}$ is given by 
\begin{align}\label{gSAVG}
\delta_{\rm gSAVG}(\bZ)=\argmin_{1 \leq j \leq J} \xi_{jd}(\bZ).
\end{align}
Observe that $\xi_{jd}$ reduces to the earlier transformation $\xi^{(0)}_{jd}$ if we consider $\gamma(t)=t$ and $\phi(t)=t$ in equation \eqref{modifyL2}. So, the gSAVG classifier is a generalization of the SAVG classifier. 

\subsection{Generalization of NN-MADD Classifier}

For a test point $\bZ \in \mathbb{R}^d$, the usual nearest neighbor (NN) classifier is defined as follows:
\begin{align}
&\delta_{\rm NN}(\bZ)=\argmin_{1 \leq j \leq J} \tau_{jd}(\bZ),
\end{align}
where $\tau_{jd}(\bZ)=\min_{\bX\in\rchi_j} \|\bZ-\bX\|$ for $1\leq j\leq J$. In high dimensions, the NN classifier perfectly classifies a new observation when $\nu^2_{jj^\prime}>|\sigma^2_j-\sigma^2_{j^\prime}|$ for all $1 \leq j\neq j^{\prime} \leq J$ \citep[see][]{HMN05}. But, when this condition is violated, this classifier may behave erratically \citep[see, e.g., ][]{PMG2016}. To avoid this problem, \cite{PMG2016} proposed an approach by modifying the distance function and defined the dissimilarity between $\bZ$ and a training observation $\bX\in\rchi$ as follows:
\vspace{-0.1cm}
\begin{equation}\label{MADD_defn}
\psi^{(0)}_{d}(\bZ,\bX)= \frac{1}{n-1}\sum\limits_{\bX^\prime\in\rchi\setminus\bX}\Big|d^{-1/2} \| \bZ-\bX^\prime \|- d^{-1/2} \| \bX-\bX^\prime \|\Big|. 
\vspace{-0.1cm}
\end{equation}
The dissimilarity $\psi^{(0)}_d$ is called the mean absolute difference of distances (MADD). The NN classifier based on MADD is defined as
\begin{align}
&\delta_{\rm NN-MADD}(\bZ)=\argmin_{1 \leq j \leq J} \tau^{(0)}_{jd}(\bZ),
\end{align}
where $\tau^{(0)}_{jd}(\bZ)=\min_{\bX\in\rchi_j} \psi^{(0)}_{d}(\bZ,\bX)$ for $1 \leq j \leq J$. The NN-MADD classifier perfectly classifies a new observation in the HDLSS setup when $\nu^2_{jj^\prime}>0$ or $\sigma_{j}^2\neq\sigma_{j^\prime}^2$ for all $1 \leq j \neq j^\prime \leq J$. This condition is clearly weaker than the one for the usual NN classifier stated above. However, this classifier too performed quite poorly in Examples~\ref{ex1}, \ref{ex2} and \ref{ex3}, where the condition was violated.

Here again, the problem lies in the use of Euclidean distance in the construction of $\psi^{(0)}_d$. To resolve this issue, we use the new distance function $h_d$ defined in \eqref{modifyL2} to modify the transformation $\psi^{(0)}_{d}$ given in \eqref{MADD_defn} as follows:
\begin{equation} \label{gMADD}
\psi^{\phi,\gamma}_d({\bZ},\bX) \equiv \psi_d({\bZ},\bX) =\frac{1}{n-1}\sum\limits_{\bX^\prime\in\rchi\setminus\bX}\big|h_d(\bZ,\bX^\prime)-h_d(\bX,\bX^\prime)\big|.
\end{equation}
The dissimilarity index $\psi_d$ is referred to as mean absolute difference of generalized distances (or, generalized MADD and hence, abbreviated as gMADD). Using gMADD, we define $\tau_{jd}(\bZ)=\min_{\bX\in\rchi_j} \psi_d(\bZ,\bX)$ for $1 \leq j \leq J$. The associated nearest neighbor classifier is defined as
\begin{align} \label{gMADD_classifier}
\delta_{\rm NN-gMADD}(\bZ)=\argmin_{1 \leq j \leq J} \tau_{jd}(\bZ).
\end{align} 

\noindent
If we consider $\gamma(t)=t$ and $\phi(t)=\sqrt{t}$ in \eqref{modifyL2}, then $\psi_d$ reduces to $\psi^{(0)}_{d}$ defined in \eqref{MADD_defn}. 
Consequently, the NN-gMADD classifier reduces to the NN-MADD classifier.

\begin{figure}[htp]
	\begin{center}
		\captionsetup{justification=centering}
		\includegraphics[width= \linewidth,height=0.32\linewidth]{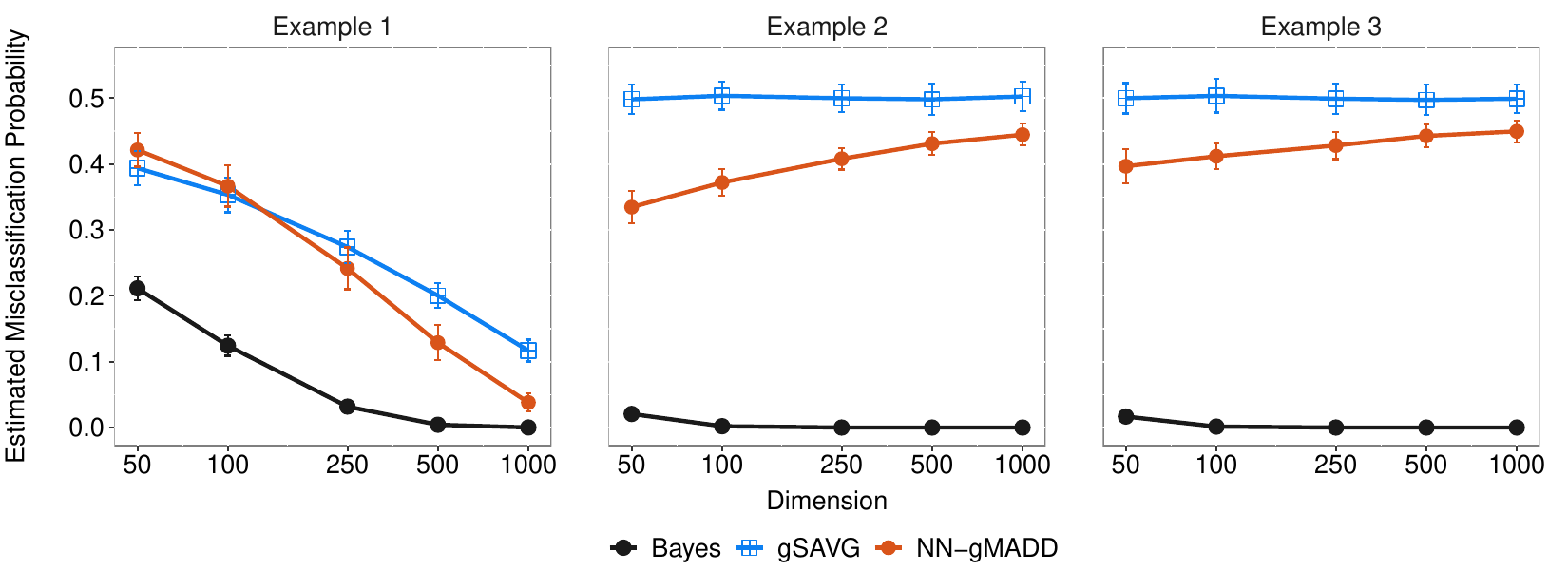}
	\end{center}
	\caption{Average misclassification rates (along with the standard errors) based on $100$ repetitions of the gSAVG and NN-gMADD classifiers are plotted for increasing values of $d$ (in logarithmic scale).}
	\label{plot1}
\end{figure}

Recall that in Examples~\ref{ex1}, \ref{ex2} and \ref{ex3} we have $\nu_{12}^2=0$ and $\sigma_1^2=\sigma_2^2$. So, both the classifiers SAVG and NN-MADD (based on Euclidean distances) performed quite poorly (see Figure~\ref{plot0}). However, Figure \ref{plot1} clearly shows the superiority of the proposed gSAVG and NN-gMADD classifiers in Example~\ref{ex1} with $\gamma(t)=1-e^{-t}$ and $\phi(t)=t$. In high dimensions, they have misclassification rates close to the Bayes risk. The misclassification rates of different NN classifiers are reported by considering a single neighbor (i.e., for $k=1$) only. We observed a similar phenomenon for other values of $k$ as well. In Figure \ref{plot1}, we further observe that both the gSAVG and NN-gMADD classifiers misclassify nearly $50$\% and $45$\% (for higher values of $d$) of the test samples in Examples~\ref{ex2} and \ref{ex3}, respectively. Interestingly, the transformation $h_d$ works favourably for Example~\ref{ex1}, while it is quite intriguing to note that it fails to yield good performance in Examples~\ref{ex2} and \ref{ex3} for high $d$. In the next subsection, we study the reason behind this behavior of the proposed classifiers in high dimensions. We begin by studying the theoretical behavior of the transformation $h_d$ in the HDLSS asymptotic regime.

\vspace{-0.2in}

\subsection{Behavior of Generalized Classifiers in HDLSS Asymptotic Regime}\label{behavegMADD}

Suppose that $\bU=(U_1,\ldots,U_d)^\top \sim \mathbf{F}_j$ and $\bV=(V_1,\ldots,V_d)^\top\sim \mathbf{F}_{j^\prime}$ are two independent $d$-dimensional random vectors. We denote the marginal distribution of the $i$-th component corresponding to the $j$-th population by $F_{j,i}$ for $1 \leq i \leq d$ and $1 \leq j \leq J$. To study the asymptotic behavior of $h_d^{\phi,\gamma}$, we make the following assumptions: 
\begin{itemize}
\item[] $(A1) \text{ There exists a constant } c_1 \text{ such that }{\rm E}\big(\gamma^2(|U_i-V_i|^2)\big) \leq c_1 <\infty\ \forall\ 1\leq i\leq d$. 
\item[] $(A2) \mathop{\sum\sum}_{1\leq i<i^\prime\leq d}{\rm Corr}\big(\gamma(|U_i-V_i|^2),\gamma(|U_{i^\prime}-V_{i^\prime}|^2)\big)=o(d^2)$. 
\end{itemize}

\noindent
It is evident that $(A1)$ is satisfied if $\gamma$ is bounded.
Assumption $(A2)$ holds if the component variables of the underlying populations are independent. However, it continues to hold even when the components are dependent, with some additional conditions on their dependence structure. For instance, in the case of sequence data, $(A2)$ holds when the sequence has the $\rho$-mixing property \citep[see, e.g.,][]{HMN05,bradley2005basic}.
Conditions similar to $(A2)$ have been considered previously for studying the high-dimensional behavior of different statistical methods \citep[see the review paper by][]{ASSYZM18}. Under assumptions $(A1)$ and $(A2)$, the high-dimensional behavior of $h_d^{\phi,\gamma}$ is given by the following lemma.

\begin{lem}\label{firstlemma}
Suppose that $\bU \sim \mathbf{F}_j$ and $\bV \sim \mathbf{F}_{j^\prime}$ are two independent random vectors satisfying assumptions $(A1)$ and $(A2)$ with $1 \le j,j^\prime \le J$, and $\phi$ is uniformly continuous. Then 
\begin{align*}
\big|h_d(\bU,\bV)-\tilde{h}_d(j,j^\prime)\big|\stackrel{P}{\to}0\text{ as }d\to\infty,
\end{align*}
where $\tilde{h}_d(j,j^\prime) \equiv \tilde{h}_d^{\phi,\gamma}(j,j^\prime)$ is defined as $\tilde h_d(j,j^\prime)=\phi[d^{-1}\sum_{i=1}^d{\rm E}\{\gamma(|U_i - V_i|^2)\}]$.
\end{lem}

\noindent
For $1 \leq j,j^{\prime} \leq J$, define the following quantities:
\begin{align*}
\tilde{\xi}_d^{\phi,\gamma}(j,j^\prime) \equiv \tilde{\xi}_d(j,j^\prime) & =\tilde{h}_d(j,j^\prime)-\frac{1}{2}\big [\tilde{h}_d(j^\prime,j^\prime)+\tilde{h}_d(j,j)\big ] \text{, and}\\
\tilde{\tau}_d^{\phi,\gamma}(j,j^\prime) \equiv \tilde{\tau}_d(j,j^\prime) & = \sum_{1\leq l\neq j^\prime\leq J} \Big [\frac{n_l}{n-1}|\tilde{h}_d(j^\prime,l)-\tilde{h}_d(j,l)|\Big ]+\frac{n_{j^\prime}-1}{n-1}|\tilde{h}_d(j^\prime,j^\prime)-\tilde{h}_d(j,j^\prime)|.
\end{align*}
As an immediate consequence of Lemma \ref{firstlemma}, we get the following result involving $\xi_{jd}(\bZ)$ (defined in \eqref{SgAVG}) and $\tau_{jd}(\bZ)$ (defined just above \eqref{gMADD_classifier}).

\begin{cor}\label{cor0}
If a test observation $\bZ \sim \mathbf{F}_j$, then for any $1\leq j^\prime\leq J$ we have
\begin{enumerate}[(a)]
\item $\big|\big\{\xi_{j^\prime d}(\bZ)-\xi_{jd}(\bZ)\big\}-\tilde{\xi}_d(j,j^\prime)\big|\stackrel{P}{\to}0\text{ as }d\to\infty,$ 
\item $\big |\{\tau_{j^\prime d}(\bZ)-\tau_{jd}(\bZ)\} -\tilde{\tau}_d(j,j^\prime)\big|\stackrel{P}{\to}0\text{ as }d\to\infty.$
\end{enumerate}
\end{cor}

\noindent
From the definition, it is clear that $\tilde{\xi}_d$ is symmetric (i.e., $\tilde{\xi}_d(j,j^\prime)=\tilde{\xi}_d(j^\prime,j)$) and $\tilde{\xi}_d(j,j)=0$ for $1 \leq j,j^{\prime} \leq J$. 
Recall that $\delta_{\rm gSAVG}$ classifies $\bZ \sim \mathbf{F}_j$ correctly if $\xi_{j^\prime d}(\bZ)-\xi_{jd}(\bZ)>0$ for all $j^\prime \neq j$. So, for good performance of gSAVG in high dimensions, it is expected that we have $\tilde{\xi}_d(j,j^\prime)>0$ for large values of $d$.
On the other hand, the constant $\tilde{\tau}_d(j,j^\prime)$ is non-negative and $\tilde{\tau}_d(j,j)=0$ for all $1\leq j,j^\prime\leq J$ by definition. 
Again, it is desirable to have $\tilde{\tau}_d(j,j^\prime)>0$ for large values of $d$, to ensure good performance of the NN-gMADD classifier. Both these requirements are met by choosing the functions $\phi$ and $\gamma$ appropriately, as stated in the following lemma.

\begin{lem}\label{thirdlemma}
Let $\gamma$ have non-constant, completely monotone derivative on $\mathbb{R}^+$. Then, the following results hold.\\
(a) If $\phi$ is concave, then $\tilde{\xi}_d(j,j^\prime)\geq 0$, and $\tilde{\xi}_d(j,j^\prime)=0$ if and only if $F_{j,i}=F_{j^\prime,i}$ for all $1 \leq i \leq d$.\\
(b) If $\phi$ is one-to-one, then $\tilde{\tau}_d(j,j^\prime)=0$ if and only if $F_{j,i}=F_{j^\prime,i}$ for all $1 \leq i \leq d$.
\end{lem}

\noindent
Functions with non-constant, completely monotone derivatives have been considered earlier in the literature \citep[see, e.g.,][]{MR0270403, BF10}.
Lemma \ref{thirdlemma} shows that for appropriate choices of $\phi$ and $\gamma$, the quantity $\tilde{\xi}_d(j,j^\prime)$ can be viewed as a measure of separation between the two population distribution functions $\mathbf{F}_j$ and $\mathbf{F}_{j^\prime}$ for $1 \leq j\neq j^\prime \leq J$. 
In fact, this quantity attains the value zero only when the two populations have identical one-dimensional marginals, and it is related to the idea of {\it energy} \citep[see][]{SR2017}. So, it is reasonable to assume the following:
\begin{itemize}
\item[] $(A3) \text{ For every   } 1 \leq j \neq j^\prime \leq J,~  \liminf \limits_{d\to\infty} \tilde{\xi}_d(j,j^\prime)>0$.
\end{itemize} 
This assumption ensures that separation among the populations is asymptotically non-negligible. A similar condition for $\tilde{\tau}_d(j,j^\prime)$ follows from assumption $(A3)$ (see Lemma~\ref{xi_implies_phi} in Appendix \ref{AppendixA}). The following theorem states the high-dimensional behavior of the proposed classifiers under these assumptions.

\begin{thm}\label{thmgSAVG}
Define $n_0=\min\{n_1,\ldots,n_J\}$. If assumptions \textrm{(A1)--(A3)} are satisfied, then\\
(a) for any $n_0 \geq 2$, the misclassification probability of the gSAVG classifier converges to zero as $d\to \infty$, and\\
(b) for any $k \le n_0$, the misclassification probability of the $k$-NN classifier based on gMADD converges to zero as $d\to \infty$.
\end{thm}

\noindent
When the underlying distributions have different marginal distributions, Theorem \ref{thmgSAVG} suggests that classifiers based on the transformation $h_d^{\phi,\gamma}$ should have excellent performance if $\phi$ and $\gamma$ are chosen appropriately. The choice $\phi(t)=t$ satisfies the conditions of Lemmas~\ref{firstlemma} and \ref{thirdlemma}.
There are several choices of $\gamma$ that satisfy the conditions stated in Lemma~\ref{thirdlemma} \citep[see][p.1338]{BF10}. In particular, $\gamma(t)=1-e^{-t}$ satisfies these conditions.

Let us now recall Figure \ref{plot1}. In Example~\ref{ex1}, the one-dimensional marginals of $\mathbf{F}_1$ are all $N(0,5/3)$, while for $\mathbf{F}_2$ the marginals are $t_5$. So, there is difference in the one-dimensional marginal distributions and assumptions $(A1)-(A3)$ are satisfied in this example. On the other hand, the marginal distributions of both classes are same (namely, $N(0,1)$) in Examples~\ref{ex2} and \ref{ex3}. As a result, assumption $(A3)$ is violated and Theorem \ref{thmgSAVG} fails to hold in these two examples. 

\section{Further Generalization Using Groups of Variables} \label{Improve}

In Figure \ref{plot1}, we have observed that the proposed classifiers fail to discriminate among populations for which the one-dimensional marginals are identical (recall Examples~\ref{ex2}~and~\ref{ex3}). However, in Example~\ref{ex2} we have information in `groups of variables' and the groups are quite prominent. If we can capture this information in the joint structure of the sub-vectors (instead of extracting information only from the $d$ univariate components) and modify our classifiers accordingly, it is expected that the classifiers will perform better. In this section, we use this idea to further generalize the transformations $\xi_d^{\phi,\gamma}$ and ${\tau}_d^{\phi,\gamma}$ so that populations can be discriminated even when the one-dimensional marginals are same. 

To build the next step of generalization, we assume that the component variables of a high-dimensional vector have an implicit property of forming groups of variables.
By groups of variables, we simply mean a non-overlapping collection of variables.
We will address the problem of finding these groups in practice later in Section \ref{cluster}. Meanwhile, let us assume that the groups are known, i.e., the components of a $d$-dimensional vector $\bu$ are partitioned into $b$ known groups. Let $\mathcal{C} = \{C_1,\ldots, C_{b}\}$ represent the collection of these groups, where $C_i=\{{l_{d_{i-1}+1}},\ldots,{l_{d_{i}}}\}$ with $d_0=0$ and $1 \leq i \leq b$. 
Now, consider the sub-vector $\bu_i=(u_{l_{d_{i-1}+1}},\ldots,u_{l_{d_i}})^\top$ of dimension $d_i$ for $1 \leq i \leq b$.
We propose a modification of $h_d^{\phi,\gamma}$ so that the discriminants can extract information from the distributions of these sub-vectors (i.e., groups of component variables). 

For two vectors $\bu = (\bu_1^\top,\ldots,\bu_b^\top)^\top$ and $\bv = (\bv_1^\top,\ldots,\bv_b^\top)^\top$, we define a generalized dissimilarity measure as follows:
\begin{align}\label{hbdef}
h_b^{\phi,\gamma}(\bu,\bv) \equiv h_b(\bu,\bv)=\phi \bigg [\frac 1 b \sum_{i=1}^b\gamma\Big(d_i^{-1} \|\bu_i - \bv_i\|^2\Big) \bigg ].
 \vspace{-0.1cm}
\end{align}
We would like to point out the notational similarity between equations \eqref{hbdef} and \eqref{modifyL2}.
Throughout the article, we use the convention that with suffix $d$, we denote the generalized distance based on component variables as defined in \eqref{modifyL2}, while with suffix $b$, we denote the generalized distance based on groups of variables as defined in \eqref{hbdef}.

We first modify the gSAVG classifier defined in \eqref{gSAVG} as follows. Using the transformation $h_b^{\phi,\gamma}$, we define
\begin{align}\label{TbgSAVG}
\xi^{\phi,\gamma}_{jb}(\bZ) \equiv \xi_{jb}(\bZ) = \frac{1}{n_j} \sum_{\bX\in\rchi_j}h_b(\bZ,\bX)-D_b(\rchi_{j}|\rchi_j)/2,
\end{align}
where $D_b(\rchi_{j}|\rchi_j)=\{n_j(n_j-1)\}^{-1} \sum_{\bX,\bX^{\prime}\in\rchi_j}\hspace{-0.1cm} h_b(\bX,\bX^{\prime})$ for $1 \leq j \leq J$. Now, the block-generalized SAVG (bgSAVG) classifier is defined as
\begin{align}\label{bgSAVG}
\delta_{\rm bgSAVG}(\bZ)=\argmin_{1 \leq j \leq J} \xi_{jb}(\bZ).
\end{align} 

Similarly, we modify the NN-gMADD classifier defined in \eqref{gMADD_classifier} as follows. Define
\begin{equation} \label{bgMADD}
\psi^{\phi,\gamma}_b({\bZ},\bX) \equiv \psi_b({\bZ},\bX) =\frac{1}{n-1}\sum\limits_{\bX^\prime\in\rchi\setminus\bX}\big|h_b(\bZ,\bX^\prime)-h_b(\bX,\bX^\prime)\big|,
\vspace{-0.1cm}
\end{equation}
and $\tau_{jb}(\bZ)=\min_{\bX\in\rchi_j} \psi_b(\bZ,\bX)$ for $1 \leq j \leq J$. The associated nearest neighbor classifier is now defined as:
\begin{align} \label{bgMADD_classifier}
\delta_{\rm NN-bgMADD}(\bZ)=\argmin_{1 \leq j \leq J} \tau_{jb}(\bZ).
\end{align}
We refer to $\delta_{\rm NN-bgMADD}$ as the NN classifier based on block-generalized MADD (or, the NN-bgMADD classifier).

\begin{figure}[H]
	\begin{center}
		\captionsetup{justification=centering}
		\includegraphics[width=\linewidth, height=0.3\linewidth]{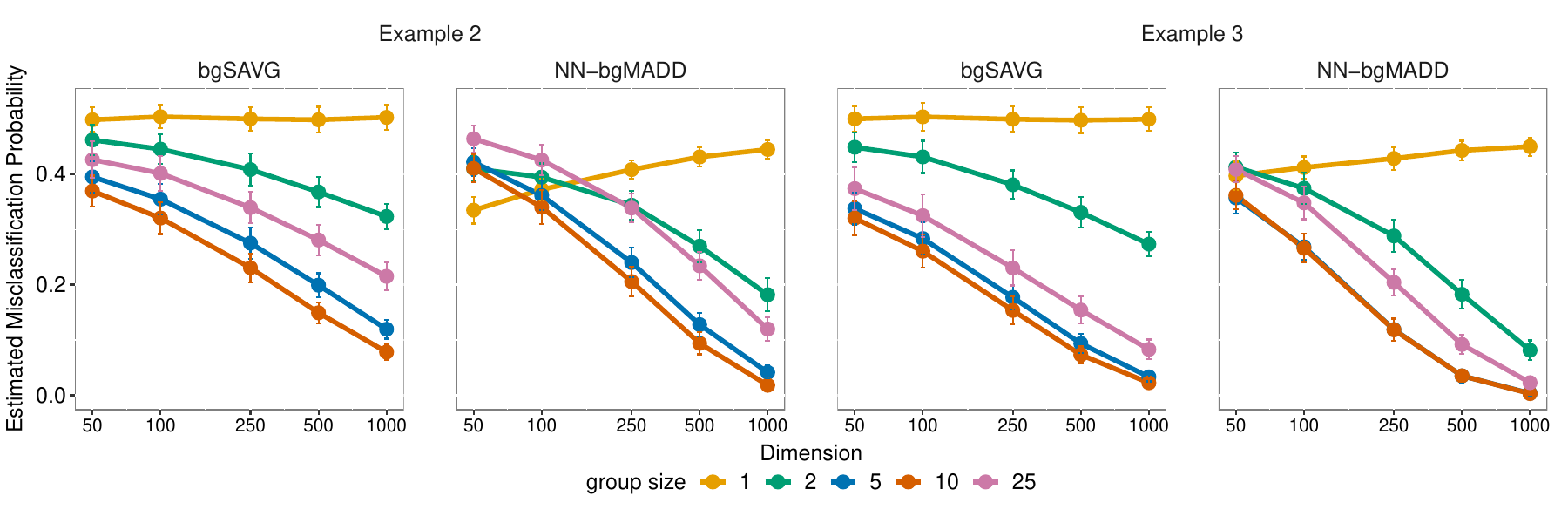}
	\end{center}
	\vspace{-0.25cm}
	\caption{Average misclassification rates (along with the standard errors) based on $100$ repetitions of the bgSAVG and NN-bgMADD classifiers are plotted with varying group sizes for increasing values of $d$ (in logarithmic scale).} 
	\label{plot3_B}
\end{figure}

Let us now investigate the performance of the proposed classifiers in Examples~\ref{ex2}~and~\ref{ex3}. The choice of groups is quite clear in Example~\ref{ex2} (we have $d_i=10$ for all $1 \leq i \leq b$ with $C_1=\{1,\ldots,{10}\}$; $C_2=\{{11},\ldots,{20}\}$; and so on), but it is not so straightforward in Example~\ref{ex3}. In both examples, we formed equal-sized groups using consecutive variables with varying choices of the group sizes, and the corresponding results are shown in Figure~\ref{plot3_B}. 

Figure \ref{plot3_B} clearly shows the superiority of the modified (both bgSAVG and NN-bgMADD with $\gamma(t)=1-e^{-t}$ and $\phi(t)=t$) classifiers when compared with the gSAVG and NN-gMADD (i.e., $d_i=1$ for all $1 \leq i \leq d$) classifiers. In high dimensions, the block-generalized classifiers have misclassification rates quite close to zero (even for low values of $d_i$ like $5$). On the other hand, the performance deteriorates when the value of $d_i$ is increased to $25$. Clearly, this reflects that the choice of group size is quite crucial for the proposed classifiers to perform well in practice. We provide details on the practical implementation of variable clustering for the block-generalized classifiers in Section \ref{cluster}. But first, we study the theoretical behavior of $h_b$ and the two associated classifiers, viz., bgSAVG and NN-bgMADD in the HDLSS asymptotic regime.

\subsection{Behavior of Block-Generalized Classifiers in HDLSS Asymptotic Regime} \label{bgMADD.HDLSS}

Recall that the HDLSS asymptotic behavior of the generalized distance $h_d$ (and associated classifiers) depend on the one-dimensional marginal distributions $F_{j,i}$ for $1 \leq i \leq d$ and $1 \leq j \leq J$. Similarly, the HDLSS asymptotic behavior of $h_b$ (and related classifiers) will be governed by the joint distributions of groups of variables. To this extent, let us assume that we have a common cluster structure $\mathcal{C}$ along all the $J$ classes, and $\mathcal C$ is known.  
For a random vector $\bU = (\bU_1^\top,\ldots,\bU_b^\top)^\top \sim \mathbf{F}_j$ partitioned according to $\mathcal C$, we denote the distribution function of $\bU_i$ by $\mathbf{F}_{j,i}$ for $1 \leq i \leq b$ and $1 \leq j \leq J$. 
To study the HDLSS asymptotic behavior of the newly proposed classifiers (viz., bgSAVG and NN-bgMADD), we restrict ourselves to the setting where 
the sizes of clusters $d_i$ remain bounded for $1 \leq i \leq b$. This assumption is formally stated below.
\begin{itemize}
\item[] $(A4) \text{ There exists a fixed positive integer }d_0\text{ such that }d_i\leq d_0 \text{ for all }1 \leq i \leq b$.
\end{itemize}

\noindent
It is clear from assumption $(A4)$ that $b\leq d=\sum_{i=1}^bd_i\leq bd_0$.  Hence, we can write `$b\to\infty$' and `$d\to\infty$' interchangeably.
Now, for $\bU = (\bU_1^\top,\ldots,\bU_b^\top)^\top \sim \mathbf{F}_j$ and $\bV = (\bV_1^\top,\ldots,\bV_b^\top)^\top \sim \mathbf{F}_{j^\prime}$ with $1 \le j,j^\prime \le J$, consider the following assumptions:
\begin{itemize}
\item[] $(A5) \text{ There exists a constant} ~c_2~ \text{such that }{\rm E}[\gamma^2\big({d_i}^{-1}\|\bU_i-\bV_i\|^2\big)]\leq c_2 ~\mbox{for all}~\\ 1 \leq i \leq b$. 
\item[] $(A6) \mathop{\sum\sum}_{1\leq i<i^\prime\leq b}{\rm Corr}\big[\gamma\big({d_i}^{-1}\|\bU_i-\bV_i\|^2\big), \gamma\big({d_{i^\prime}}^{-1}\|\bU_{i^\prime}-\bV_{i^\prime}\|^2 \big)\big]=o(b^2)$. 
\end{itemize}

\noindent
Assumptions $(A5)$ and $(A6)$ are generalizations of assumptions $(A1)$ and $(A2)$, respectively. As we observed earlier, choosing $\gamma$ to be bounded is sufficient to satisfy assumption $(A5)$, while assumption $(A6)$ imposes some restrictions on the dependence structure among the sub-vectors. If the sub-vectors are mutually independent, then assumption $(A6)$ is clearly satisfied. When the sub-vectors are dependent, additional conditions like weak dependence among the groups of variables are required. 
In particular, if the sequence $\{\gamma\big({d_i}^{-1}\|\bU_i-\bV_i\|^2\big), i \geq 1\}$ has the $\rho$-mixing property, then assumption $(A6)$ holds. 
A sufficient condition for $\{\gamma\big({d_i}^{-1}\|\bU_i-\bV_i\|^2\big), i \geq 1\}$ to be a $\rho$-mixing sequence is to have the sequences $\bU$ and $\bV$ to satisfy the $\rho$-mixing property (see Lemma \ref{mixinglemma} in Appendix \ref{AppendixA}). With these assumptions, we are now ready to state the high-dimensional behavior of $h_b^{\phi,\gamma}$.

\begin{lem}\label{secondlemma}
Suppose that $\bU \sim \mathbf{F}_j$ and $\bV \sim \mathbf{F}_{j^\prime}$ $(1 \leq j,j^{\prime} \leq J)$ are two independent random vectors satisfying assumptions $(A5)$ and $(A6)$. Additionally, if assumption $(A4)$ is satisfied and $\phi$ is uniformly continuous, then 
\begin{align*}
\big|h_b(\bU,\bV)-\tilde{h}_b(j,j^\prime)\big|\stackrel{P}{\to}0\text{ as }b\to\infty,
\end{align*}
where $\tilde{h}_b(j,j^\prime) \equiv \tilde{h}_b^{\phi,\gamma}(j,j^\prime)=\phi \big [ b^{-1}\sum_{i=1}^b {\rm E}\{\gamma(d_i^{-1}\|\bU_i - \bV_i\|^2)\} \big ]$.
\end{lem}

\noindent
The next result involves $\xi_{jb}(\bZ)$ (defined in \eqref{TbgSAVG}) and $\tau_{jb}(\bZ)$ (defined just above \eqref{bgMADD_classifier}), and it is a straightforward extension of Corollary \ref{cor0}. 

\begin{cor}\label{cor2}
If a test observation $\bZ\sim \mathbf{F}_j$, then for any $1\leq j^\prime\leq J$, we have
\begin{enumerate}[(a)]
\item $\big|\big\{\xi_{j^\prime b}(\bZ)-\xi_{jb}(\bZ)\big\}-\tilde{\xi}_b(j,j^\prime)\big|\stackrel{P}{\to}0\text{ as }b\to\infty,$
\item $\big |\{\tau_{j^\prime b}(\bZ)-\tau_{jb}(\bZ)\} -\tilde{\tau}_b(j,j^\prime)\big|\stackrel{P}{\to}0\text{ as }b\to\infty,$
\end{enumerate}
where, for $1 \leq j,j^{\prime} \leq J$,
$$\tilde{\xi}_b(j,j^\prime) \equiv \tilde{\xi}_b^{\phi,\gamma}(j,j^\prime)=\tilde{h}_b(j,j^\prime)-\frac{1}{2}\big [\tilde{h}_b(j^\prime,j^\prime)+\tilde{h}_b(j,j) \big ],~ and$$ 
$$\tilde{\tau}_b^{\phi,\gamma}(j,j^\prime) \equiv \tilde{\tau}_b(j,j^\prime) = \sum_{1 \leq l \neq j^\prime \leq J} \bigg [\frac{n_l}{n-1}|\tilde{h}_b(j^\prime,l)-\tilde{h}_b(j,l)|\bigg ]+\frac{n_{j^\prime}-1}{n-1}|\tilde{h}_b(j^\prime,j^\prime)-\tilde{h}_b(j,j^\prime)|.$$
\end{cor}

\noindent
Similar to the constants $\tilde{\xi}_d(j,j^\prime)$ and $\tilde{\tau}_d(j,j^\prime)$, both $\tilde{\xi}_b(j,j^\prime)$ and $\tilde{\tau}_b(j,j^\prime)$ are measures of separability between $\mathbf{F}_j$ and $\mathbf{F}_{j^\prime}$ for $1 \leq j,j^{\prime} \leq J$. 
While $\tilde{\tau}_b(j,j^\prime)$ is non-negative by definition, the same is true for $\tilde{\xi}_b(j,j^\prime)$ if $\phi$ is concave. Moreover, under conditions similar to Lemma~\ref{thirdlemma}, both $\tilde{\xi}_d(j,j^\prime)$ and $\tilde{\tau}_d(j,j^\prime)$ are strictly positive whenever $\mathbf F_j$ and $\mathbf F_{j^\prime}$ have different group distributions (i.e., $\mathbf F_{j,i} \ne \mathbf F_{j^\prime,i}$ for some $1 \le i \le b$). This is shown in the following lemma.

\begin{lem}\label{fourthlemma}
Let $\gamma$ have non-constant, completely monotone derivative on $\mathbb{R}^+$. Then, the following results hold. \\
(a) If $\phi$ is concave, then  $\tilde{\xi}_b(j,j^\prime)\geq 0$ for all $1 \leq j,j^\prime \leq J$. Moreover, $\tilde{\xi}_b(j,j^\prime)=0$ if and only if $\mathbf{F}_{j,i}=\mathbf{F}_{j^\prime,i}$ for all $1 \leq i \leq b$.\\ 
(b) If $\phi$ is one-to-one, then $\tilde{\tau}_b(j,j^\prime) = 0$ if and only if $\mathbf{F}_{j,i}=\mathbf{F}_{j^\prime,i}$ for all $1 \leq i \leq b$.
\end{lem}

To derive HDLSS asymptotic results, we require the competing populations to be asymptotically separable. So, we assume the following:
\begin{itemize}
\item[] $(A7) \text{ for every   }1\leq j\neq j^\prime\leq J,~  \liminf\limits_{b\to\infty} \tilde{\xi}_b(j,j^\prime)>0$.
\end{itemize}
This assumption ensures that separation induced by the blocks is asymptotically non-negligible. It further implies that a similar condition holds for $\tilde{\tau}_b(j,j^\prime)$ (see Lemma \ref{xi_implies_phi} in Appendix \ref{AppendixA}). 
Following our discussion preceding Lemma \ref{fourthlemma}, assumption $(A7)$ is a generalization of assumption $(A3)$ because if we have difference in the marginal distributions, then the joint distributions are bound to be different. But, the converse is clearly not true.  In other words, if two distributions $\mathbf{F}_j$ and $\mathbf{F}_{j^\prime}$ are not separable in terms of $\tilde{\xi}_b$ (respectively, $\tilde{\tau}_b$), then they are not separable in terms of $\tilde{\xi}_d$ (respectively, $\tilde{\tau}_d$).
The following theorem shows the high-dimensional behavior of the bgSAVG and NN-bgMADD classifiers under assumption $(A7)$.

\begin{thm}\label{thm_bgMADD_bgSAVG}
Define $n_0=\min\{n_1,\ldots,n_J\}$. If assumptions {\textrm (A4)--(A7)} are satisfied, then\\
(a) for $n_0 \geq 2$, the misclassification probability of the bgSAVG classifier converges to zero as $b\to \infty$,\\
(b) for any $k \le n_0$, the misclassification probability of the $k$-NN classifier based on bgMADD converges to zero as $b\to \infty$.
\end{thm}

Recall that in Examples~\ref{ex2} and \ref{ex3} we have identical marginal distributions (namely, $N(0,1)$) for both the classes, but differences in their joint distributions. Theorem \ref{thm_bgMADD_bgSAVG} states that if this information from the joint distributions can be captured by appropriately identifying the groups, then the misclassification probability for both the classifiers should decrease to $0$ as $d$ (equivalently, $b$) increases. We have already observed this in Figure \ref{plot3_B}. 

\subsection{Comparison between bgSAVG and NN-bgMADD} \label{bgSAVG_and_NN-bgMADD}

In the previous sub-section, we have observed that both bgSAVG and NN-bgMADD classifiers achieve {\it perfect classification} in high dimensions under similar conditions. But, their relative performance may vary, especially when the dimension is not sufficiently large. To demonstrate the relative behavior of these two classifiers, we now consider two examples. The first example is Example~\ref{ex2} from Section~\ref{Intro}. As a second example, we use the following.
\begin{ex}\label{ex4}
We consider two populations, where the $d$ component variables are i.i.d. For the first population, the component distribution is Cauchy with location parameter $0$ and scale $1$ (standard Cauchy), while it is Cauchy with location parameter $0.75$ and scale $0.75$ for the second one. In this example, we take $n_1=50$ and $n_2=25$ to form the training set.
\end{ex}

Let us now look into the numerical performance of the proposed classifiers in Examples~\ref{ex2} and \ref{ex4}. We keep all other parameters (e.g., the number of iterations, test sample size) associated with this simulation same as before, and set $d_i=10$ (respectively, $d_i=1$) for all $1 \leq i \leq b$ in Example~\ref{ex2} (respectively, Example~\ref{ex4}). 

\begin{figure}[htp]
\begin{center}
\captionsetup{justification=centering}
\includegraphics[height=0.325\linewidth]{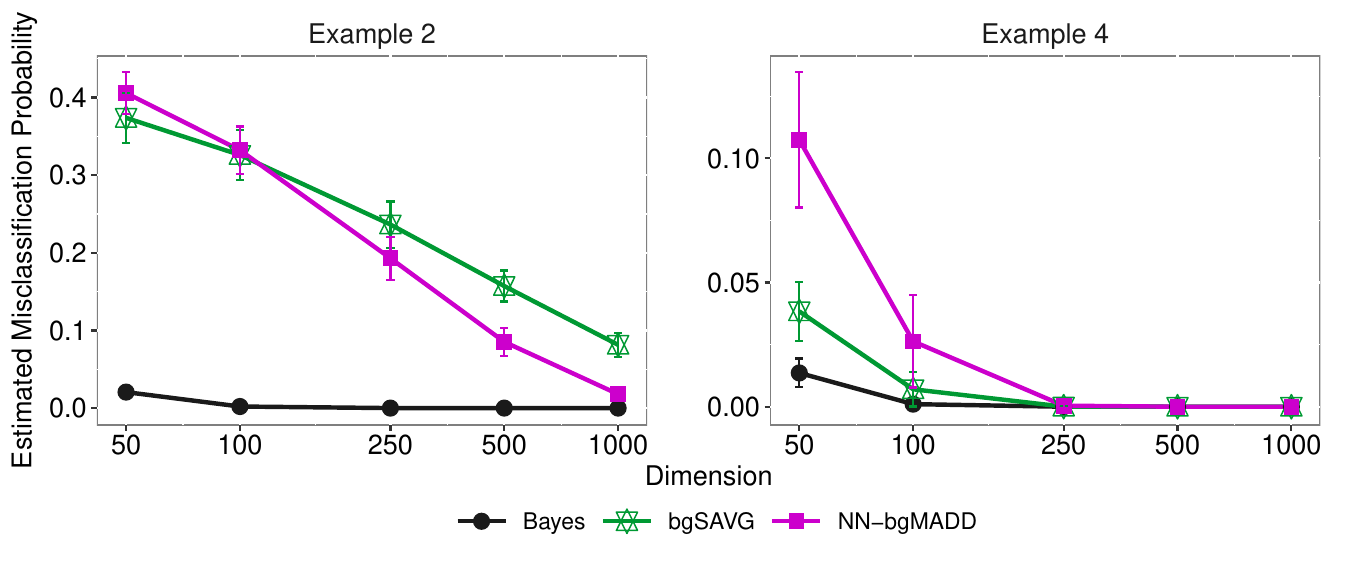}
\end{center}
\caption{Average misclassification rates (along with the standard errors) based on $100$ repetitions for the bgSAVG and NN-bgMADD classifiers are plotted for increasing values of $d$ (in logarithmic scale) for Examples~\ref{ex2} and \ref{ex4}.}
\label{plot2}
\end{figure} 

Figure \ref{plot2} clearly shows that the estimated misclassification probabilities 
for the proposed classifiers (with $\gamma(t)=1-e^{-t}$ and $\phi(t)=t$) go to $0$ with increasing values of $d$, and hence quite close to the estimated Bayes risks in Examples~\ref{ex2} and \ref{ex4}. Clearly, assumptions $(A4)-(A7)$ hold in both these examples (with bounded $\gamma$ for Example~\ref{ex4}). In Example~\ref{ex2}, the block distributions are $10$-dimensional multivariate Gaussian with different correlation structures for the two classes. The marginal distributions are Cauchy (i.e., heavy-tailed) in Example~\ref{ex4} with differences in their locations and scales. So, assumptions $(A5)$ and $(A6)$ hold with a bounded $\gamma$ function. Interestingly, bgSAVG and NN-bgMADD behave differently in these examples with one dominating the other in the respective examples.

Let us now study this phenomena in further detail. 
From the proof of Theorem \ref{thm_bgMADD_bgSAVG}, one can observe that the high-dimensional behavior of the bgSAVG and NN-bgMADD classifiers depend on the behavior of the constants $\tilde{\xi}_b(j,j^\prime)$ and $\tilde{\tau}_b(j,j^\prime)$, respectively, for $1 \leq j,j^\prime \leq J$. Consequently, the difference between these two classifiers lies in the difference between these constants. To compare between these two classifiers, we make the following assumption, which implies that the difference between $\tilde{\xi}_b(j,j^\prime)$ and $\tilde{\tau}_b(j,j^\prime)$ does not vanish as the data dimension increases.
\begin{itemize}
\item[] $(A8) \lim\inf_b |\tilde{\xi}_b(j,j^\prime)-\tilde{\tau}_b(j,j^\prime)|>0\text{ for all } 1 \leq j\neq j^\prime \leq J$.
\end{itemize}
The next theorem states the condition under which one classifier dominates the other, and vice-versa. Define the misclassification probabilities as $\Delta_{\rm bgSAVG}=\Pr[\delta_{\rm bgSAVG}({\bf X}) \neq Y]$ and $\Delta_{\rm NN-bgMADD}=\Pr[\delta_{\rm NN-bgMADD}({\bf X}) \neq Y]$, where $Y$ denotes the class label of ${\bf X}$.

\begin{thm}\label{cmpr_AVG_NN}
If assumptions $(A4)-(A6)$ and $(A8)$ are satisfied, and there exists an integer $B_1$ such that $\tilde{\xi}_b(j,j^\prime) > \tilde{\tau}_b(j,j^\prime)$ for all $b\geq B_1$ and $1\leq j\neq j^\prime \leq J$, then there exists an integer $B_2$ such that
$$\Delta_{\rm bgSAVG} \leq \Delta_{\rm NN-bgMADD} \text{ for all } b\geq B_2.$$
\end{thm}

\begin{rmk}
If the constants $\tilde{\xi}_b(j,j^\prime)$ and $\tilde{\tau}_b(j,j^\prime)$ are interchanged in the inequality (stated above), then the ordering of the misclassification probability of the respective classifiers is reversed.
\end{rmk}

\noindent
We now elaborate on this theorem for two-class problems. Recall the expressions for $\tilde{\xi}_b(1,2)$ and $\tilde{\tau}_b(1,2)$ from Corollary \ref{cor2}. The ordering between $\tilde{\xi}_b(1,2)$ and $\tilde{\tau}_b(1,2)$ clearly depend on the relationship between the constants $\tilde{h}_b(1,2)$, $\tilde{h}_b(1,1)$ and $\tilde{h}_b(2,2)$ (recall the definition from Lemma \ref{secondlemma}), and the sample sizes $n_1$ and $n_2$. A detailed case by case study on this inequality is provided by Lemma \ref{xi_better_than_psi} in Appendix \ref{AppendixA}. To draw a comparison, let us now look back at Examples~\ref{ex2} and \ref{ex4}. Clearly, the constants $\tilde{\xi}_b(1,2)$, $\tilde{\tau}_b(1,2)$ and $\tilde{\tau}_b(2,1)$ are free of $b$ in both these examples. Calculating the constants involve computing univariate/multivariate integrals. More details on these calculations can be found in Section 2 of the Supplementary. The constants take the values $\tilde{\xi}_b(1,2)=0.0101$, $\tilde{\tau}_b(1,2)=0.0470$ and $\tilde{\tau}_b(2,1)=0.0472$ in Example~\ref{ex2}, while in Example~\ref{ex4} they are $\tilde{\xi}_b(1,2)=0.0327$, $\tilde{\tau}_b(1,2)=0.0213$ and $\tilde{\tau}_b(2,1)=0.0222$ (also see Table \ref{constants_SAVG_NN}). Clearly, the value of $\tilde{\xi}_b(1,2)$ is smaller than those of $\tilde{\tau}_b(1,2)$ and $\tilde{\tau}_b(2,1)$ in Example~\ref{ex2}. Theorem \ref{cmpr_AVG_NN} suggests that the misclassification probability of the NN-bgMADD classifier should be smaller than the bgSAVG classifier for large values of $b$. This can be observed in the left panel of Figure \ref{plot2} for dimension higher than $100$. On the other hand, in Example~\ref{ex4}, the value of $\tilde{\xi}_b(1,2)$ is larger than those of $\tilde{\tau}_b(1,2)$ and $\tilde{\tau}_b(2,1)$, and one observes a role reversal in the right panel of Figure \ref{plot2}. This analysis has been continued for all the examples discussed in this article later in Section \ref{sim}. 

A few words are called for assumption $(A8)$, which holds under various scenarios. In particular, if the component variables of the underlying distributions are i.i.d., then $\tilde{\xi}_b$ and $\tilde{\tau}_b$ are free of $b$. 
Some more general conditions are discussed in Lemma \ref{xi_better_than_psi} of Appendix \ref{AppendixA}. It can also be shown that assumption $(A8)$ holds under more general cases like Example~\ref{ex2} (see Remark \ref{A8_const} in Appendix \ref{AppendixA}).

\section{Practical Implementation of Variable Clustering} \label{cluster}

For practical implementation of the methodology defined in the previous section, we need to find an appropriate clustering $\mathcal C$ of the component variables. The basic idea is to partition a $d$-dimensional vector $\bU$ into $b$ disjoint groups (or, sub-vectors) $\bU_1,\ldots,\bU_b$ such that the variables in the same sub-vector are more \emph{similar} to each other than the variables in different sub-vectors. 
Such phenomena (groups of variables) arises naturally in scientific areas like genomics. In microarray gene expressions, genes that share similar pattern of expression are usually put into a cluster \citep[see, e.g.,][]{eisen1998cluster}, while such groups of variables also play a key role in bio-diversity modeling \citep[see, e.g.,][]{faith1996environmental}. 

We would like to emphasize that the order in which the component variables are arranged in a sub-vector is irrelevant in this context. Therefore, we use the terms `group' and `sub-vector' interchangeably. 
Here, we assume the same grouping of component variables for all $J$ populations. In general, different populations may have different groups of component variables. But, in a two-class problem, if the group structure of one population is either finer (or, coarser) w.r.t. the other population, then we can assume the coarser structure for both the populations. For more than two classes, if the group structure of one population is coarser than all the competing populations, it is sufficient to use the coarsest structure across all populations. 
In any case, our problem is essentially that of clustering $d$ variables with $n$ observations for each variable (i.e., $d$ observations in $\mathbb{R}^n$). Any appropriate clustering algorithm \citep[see, e.g.,][]{hastie2009elements} can be used for this purpose. 
To summarize, one can view this idea of constructing groups as a problem of clustering the component variables using an appropriate measure of similarity. 
So first, let us discuss the idea of \emph{similarity} (equivalently, {\it dissimilarity}) among variables.

For the HDLSS asymptotic results, we need variables from different groups (or, clusters) to have weak dependence (see assumption $(A6)$). On the other hand, highly dependent variables are natural candidates to be included in the same cluster. A reasonable measure of dependence between two components is the absolute value of their correlation coefficient. Let $r(i,i^{\prime})$ denote the correlation between the $i$-th and the $i^{\prime}$-th components for $1 \leq i,i^\prime \leq d$. If $|r(i,i^{\prime})|$ is high, then we say that the $i$-th and the $i^{\prime}$-th components are strongly associated, or `similar'. While $|r(i,i^{\prime})|$ is a measure of similarity, $1-|r(i,i^{\prime})|$ can be considered as a measure of {\it dissimilarity}. We use the agglomerative hierarchical clustering algorithm with average linkage \citep[see, e.g.,][]{hastie2009elements} and $1-|r(i,i^{\prime})|$ as the pairwise dissimilarity measure to obtain clusters of components. Starting with each component variable as a single cluster, hierarchical methods merge the least dissimilar clusters in turn until all the components are put together in one single cluster. For heavy-tailed distributions (like the Cauchy distribution), a robust measure of correlation can be used.

In hierarchical clustering, each level in the hierarchy induces a set of clusters, and the whole hierarchy (visualized as a dendrogram) represents a nested structure among the clusters obtained at different levels (see Figure \ref{dendo} below). The height of each level represents the dissimilarity between the clusters that are merged together at that level. In other words, each cluster structure is represented by the height of the level corresponding to that structure. Therefore, finding an appropriate clustering is equivalent to identifying a suitable level in the hierarchy. Suppose $\bH$ is the set of all heights that are obtained at different levels of clustering. We order the values in $\bH$, and find the $\alpha$-th percentile $H_{\alpha}$ for different values of $\alpha \in A=\{0,0.1,\ldots,0.9,1\}$. 
For each fixed $\alpha$, we obtain a clustering induced by $H_{\alpha}$. Note that the number of clusters is non-increasing in $\alpha$, while the size of each cluster is non-decreasing. In particular, $H_0$ corresponds to the case where each cluster consists of a single component variable only, i.e., $b=d$. On the other hand, $H_1$ leads to the clustering where all the $d$ components are put together in a single cluster.

\begin{figure}[H]
	\vspace{-0.1in}
	\centering
	\subfloat{
		\includegraphics[width= 0.8\linewidth,height=0.35\linewidth]{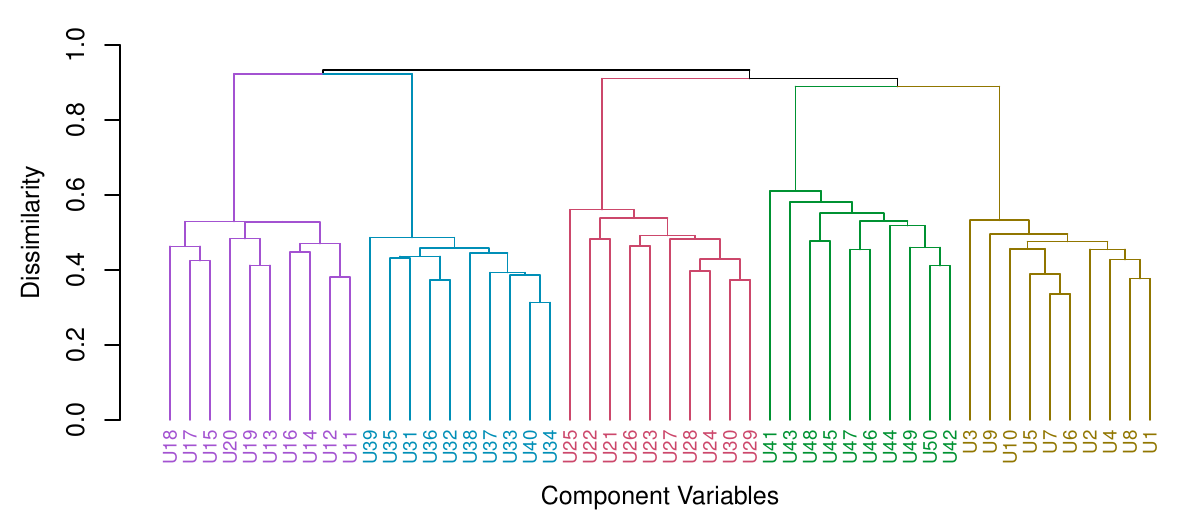}
	}
	\caption{Dendrogram showing structures of clusters in Example~\ref{ex2} for one run of a simulation with $d=50$.}
	\label{dendo}
\end{figure}

We demonstrate this idea using Example~\ref{ex2}. In this example (with $d=50$ and $n_1=n_2=50$), the groups of component variables (common across both classes) are the sets $C_1=\{1,\ldots,{10}\}; C_2=\{{11},\ldots,{20}\};\ldots; C_5=\{{41},\ldots,{50}\}$. 
We consider a simulated realization from this example. Figure \ref{dendo} shows the dendrogram for this data. At $H_{0.9}=0.67$, we obtain five clusters in Figure \ref{dendo}. The distinct clusters are indicated with five different colors, while the components corresponding to each cluster are marked with the same color in Figure \ref{dendo}.
Clearly, the method correctly assigns desired components to the respective groups (up to a permutation of the components within each group). Once the groups $\bU_1,\ldots,\bU_b$ have been identified, we can compute $h_b^{\phi,\gamma}$ as in equation \eqref{hbdef} and classify observations using the bgSAVG classifier, or the NN-bgMADD classifier introduced in Section \ref{Improve}.

\begin{figure}[htpb]
	\vspace{0.5cm}
	\begin{center}
		\captionsetup{justification=centering}
		\includegraphics[width= 0.95\linewidth, height=0.315\linewidth]{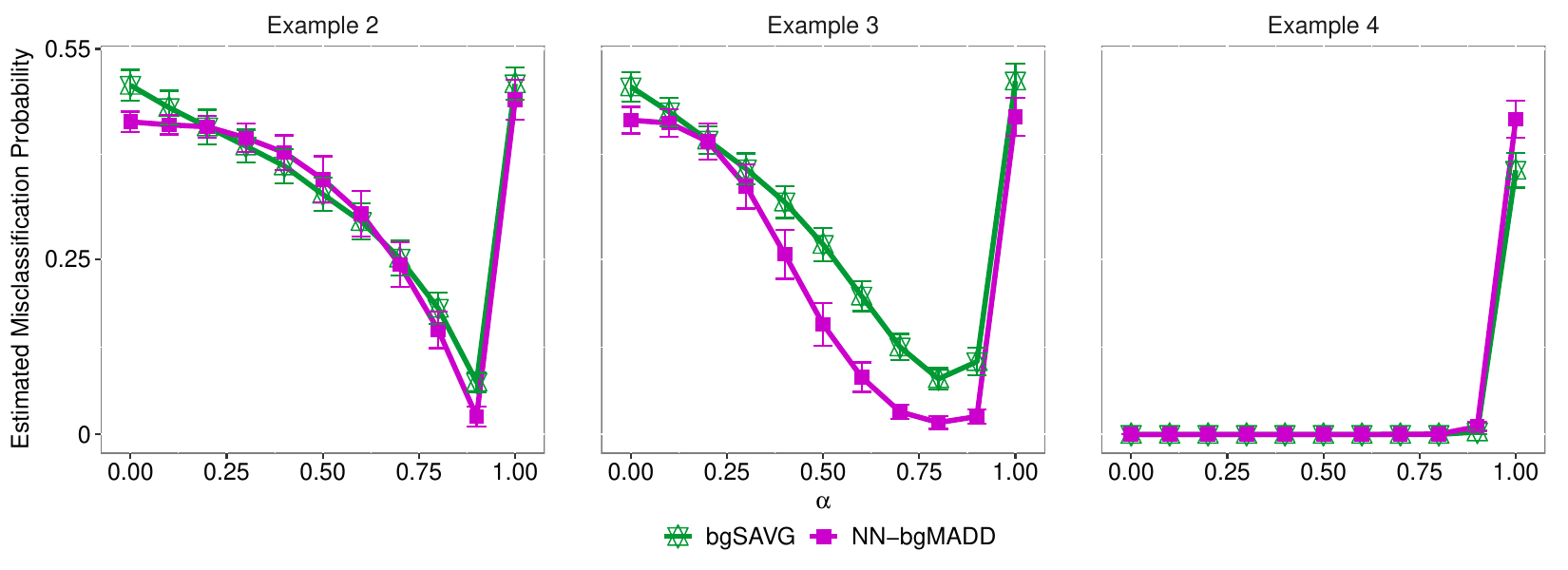}
	\end{center}
	\caption{Average misclassification rates (along with the standard errors) based on $100$ repetitions of the bgSAVG and NN-bgMADD classifiers for increasing values of $\alpha$ in Examples~\ref{ex2}--\ref{ex4}.}
	\label{plot_3}
\end{figure}

It is evident from Figure \ref{dendo} that the choice of $H_{\alpha}$ (or, equivalently $\alpha$) is crucial in finding the `true' cluster structure. However, our task here is not to find the `true' cluster structure in the variables, but rather to find cluster structures that are useful for classification. Similar to the cluster structure, the performance of a classifier should also depend on the choice of $\alpha$. To investigate this, we looked at the misclassification rates of the bgSAVG and the NN-bgMADD classifiers (with  $\gamma(t)=1-e^{-t}$ and $\phi(t)=t$) in Examples~\ref{ex2}--\ref{ex4} for varying choices of $\alpha$ (which corresponds to different cluster structures). Clearly, Figure \ref{plot_3} shows that the classification performance depends crucially on the choice of $\alpha$.

To obtain a data driven choice of $\alpha$, we use the idea of leave-one-out cross-validation method \citep[see, e.g.,][]{hastie2009elements}. For a fixed value of $\alpha \in A$, define 
$$e_{\alpha}= \frac{1}{n} \sum_{i=1}^n \mathbb{I}\{\delta_{\alpha}^{-i}(\bX_i) \neq Y_i\}.$$
Here, $\delta_{\alpha}^{-i}$ is a classifier (bgSAVG or NN-bgMADD) constructed by leaving out the $i$-th sample from the training data for $1 \leq i \leq n$. 
Define $\hat \alpha = \arg\min_{\alpha \in A}{e_{\alpha}}$. We use the clustering induced by $H_{\hat \alpha}$ as the optimal one to carry out further analysis.

\begin{figure}[!h]
\begin{center}
		\captionsetup{justification=centering}
		\vspace*{-0.15in}
		\hspace*{-0.2in}
		\subfloat{
			\includegraphics[width= \linewidth,height=0.25\linewidth]{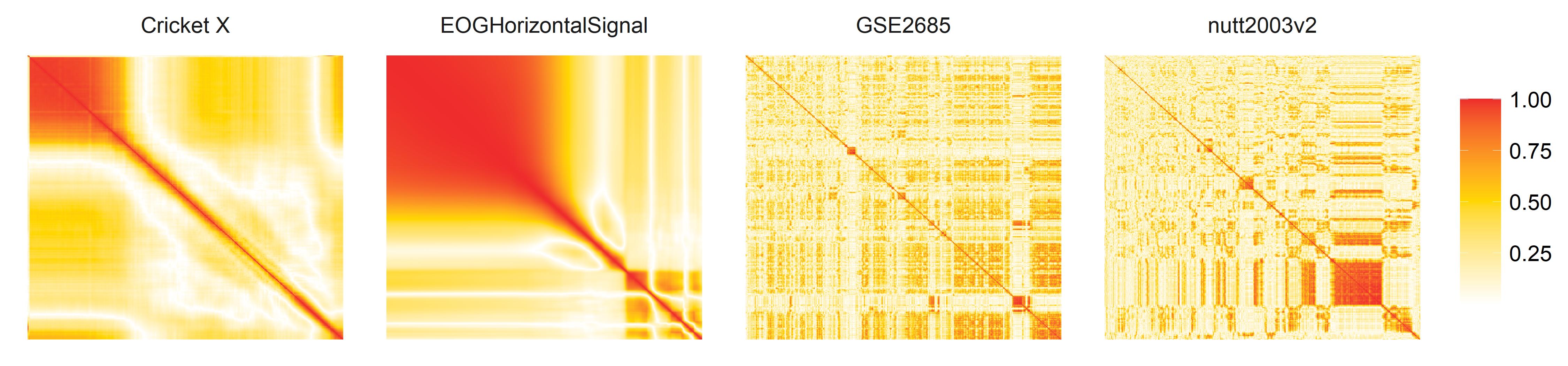} 
		}
\end{center}
\caption{Absolute of sample correlation matrices for the four benchmark data sets.} 
\vspace*{-0.1in}
\label{real_motivation}
\end{figure}

As we already mentioned, the idea of grouping in component variables can be found in several real data scenarios as well. To realize this, we plot similarity matrices of the components for {\tt four} high-dimensional data sets from {\tt three} different data archives. 
The {\tt Cricket X} and {\tt EOGHorizontalSignal} data sets are both $12$ class problems from the UCR Time Series Classification Archive \citep[see][]{UCRArchive2018} with $(n_j,d)$ as $(32,300)$ and $(30,1250)$ for $1 \leq j \leq J$. The first data is related to motion, while the second data set was collected from an electro-oculography (EOG). In Figure \ref{real_motivation}, we distinctly observe about $1$ group and $2$ groups (the second group has some smaller blocks) for these two data sets, respectively. 
The {\tt GSE2685} data set (available at the Microarray database: \url{http://www.biolab.si/supp/bi-cancer/projections/}) comprises of gene expression measurements of $30$ tissue samples distributed over $3$ classes ($8$ normal gastric tissues, $5$ diffuse gastric tumors and $17$ intestinal gastric tumors). The blocks are unclear if we plot all $4522$ genes (variables) in this data set, so we have created a plot with reduced number of (about $1500$) variables.  
In the {\tt nutt2003v2} data set (available at the Compcancer database: \url{https://schlieplab.org/Static/Supplements/CompCancer/datasets.htm}), it was investigated whether gene expression profiling could be used to classify high-grade gliomas. Microarray analysis was used to determine the expression of approximately $12000$ genes in a set of $28$ glioblastomas which were classified as classic (C), or non-classic (N). The plots in Figure \ref{real_motivation} also indicate the presence of group structure in these two gene expression data sets. We give a more detailed analysis of these four real data sets later in Section \ref{real}.

\section{Simulation Studies} \label{sim}

In this section, we thoroughly analyze some high-dimensional simulated data sets to compare the performance of the classifiers proposed in Sections \ref{general} and \ref{Improve}. We have already introduced Examples~\ref{ex1}--\ref{ex3} in Section \ref{Intro}, and Example~\ref{ex4} in Section \ref{Improve}. Four new examples are considered in this section to demonstrate the performance of the proposed classifiers.
\begin{ex}\label{ex5}
The two distributions are $N_d(\0_d,\bI_d)$ and $N_d(0.25\1_d,\bI_d)$, where $\0_d$ is the $d$-dimensional vector of zeros, $\1_d$ is the $d$-dimensional vector of ones and $\bI_d$ is the $d \times d$ identity matrix. Note that the $d$ component variables are i.i.d.\ for both the populations.
\end{ex}
\begin{ex}\label{ex6}
We again consider two Gaussian distributions $N_d(\0_d,\bI_d)$ and $N_d(\0_d,0.5\bI_d)$. Here, the $d$ component variables are i.i.d.\ similar to Example~\ref{ex5}.
\end{ex}
\begin{ex}\label{ex7}
The distributions are $\mathbf{F}_1(\bu) = \prod_{i=1}^2 \mathbf{F}_{1,i}(\bu_i)$ and $\mathbf{F}_2(\bu) = \prod_{i=1}^2 \mathbf{F}_{2,i}(\bu_i)$, with $\mathbf{F}_{1,1} \equiv N_{\lfloor\frac{d}{2}\rfloor}(\0_{\lfloor\frac{d}{2}\rfloor},\bI_{\lfloor\frac{d}{2}\rfloor})$, $\mathbf{F}_{1,2} \equiv N_{d-\lfloor\frac{d}{2}\rfloor}(\0_{d-\lfloor\frac{d}{2}\rfloor},0.5\bI_{d-\lfloor\frac{d}{2}\rfloor})$, $\mathbf{F}_{2,1} \equiv N_{\lfloor\frac{d}{2}\rfloor}(\0_{\lfloor\frac{d}{2}\rfloor}, 0.5\bI_{\lfloor\frac{d}{2}\rfloor})$ and $\mathbf{F}_{2,2} \equiv N_{d-\lfloor\frac{d}{2}\rfloor}(\0_{d-\lfloor\frac{d}{2}\rfloor},\bI_{d-\lfloor\frac{d}{2}\rfloor})$. Here, $\lfloor \cdot \rfloor$ denotes the floor function.
\end{ex}
\begin{ex}\label{ex8} We take $\mathbf{F}_1\equiv N_d(\0_d,\bI_d)$ and $\mathbf{F}_2(\bu) = \prod_{i=1}^b \mathbf{F}_{2,i}(\bu_i)$, with $\mathbf{F}_{2,i} \equiv PN_{10}(\1_{10},10)$ for all $1 \leq i \leq b$. Here, $PN_{10}(\bbeta,\alpha)$ denotes the ten-dimensional multivariate power normal distribution with parameters $\bbeta=(\beta_1,\ldots,\beta_{10})^{\top}$ with $\beta_i>0$ for all $1 \leq i \leq 10$ and $\alpha>0$ \cite[see, e.g.,][]{KG2013}. Note that $\beta_i=1$ for all $1\leq i\leq 10$ implies that the one-dimensional marginals of $\mathbf{F}_2$ are all standard normal.
\end{ex}

\begin{figure}[htp]
	\vspace*{-0.3in}
	\begin{center}
		\captionsetup{justification=centering}
		\includegraphics[width=\linewidth, height=1.4\linewidth]{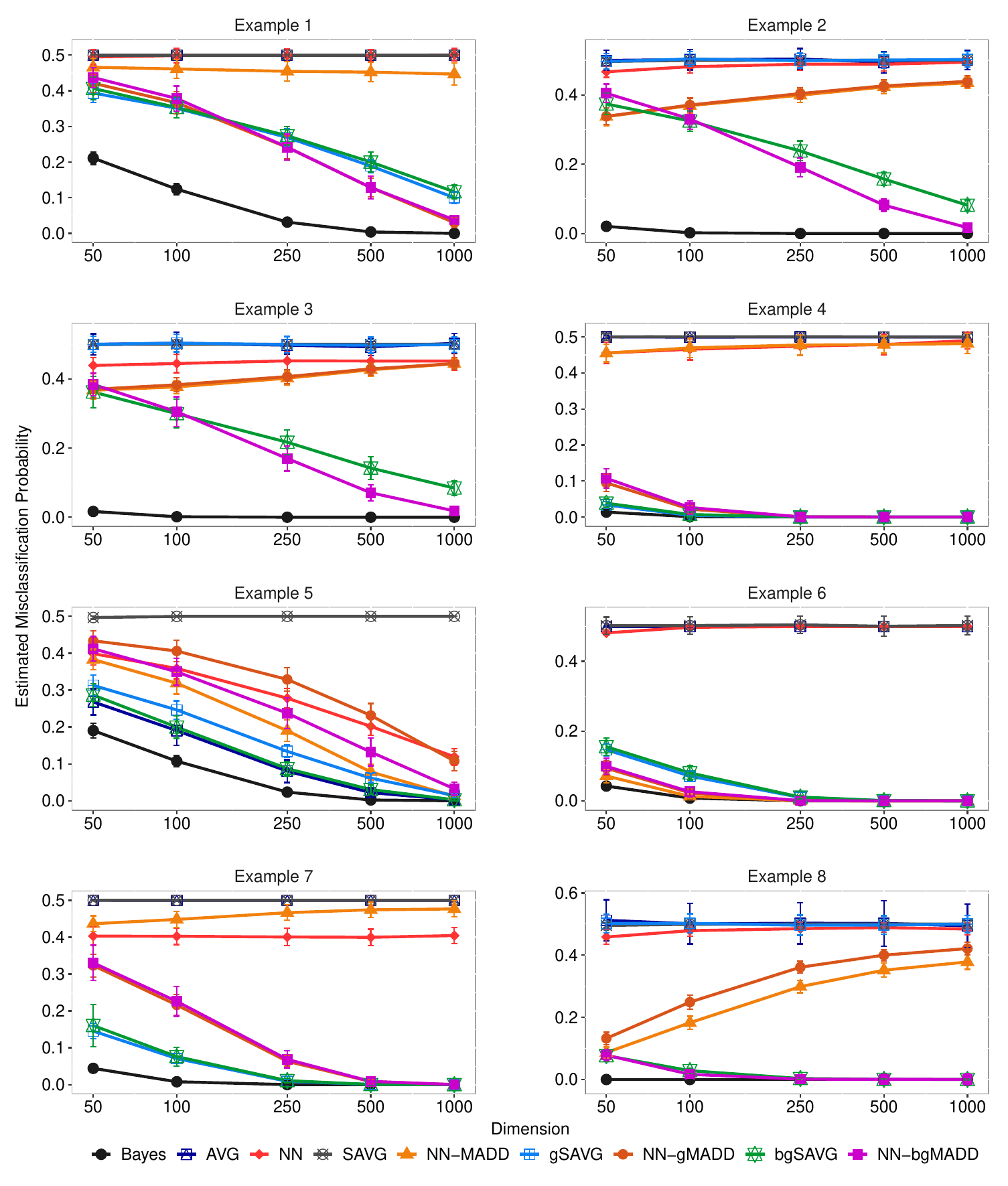}
	\end{center}
	\caption{Average misclassification rates (along with the standard errors) based on $100$ repetitions for different classifiers are plotted for increasing values of $d$ (in logarithmic scale).} 
	\label{plot3_a}
\end{figure}

In each example, we simulated data for $d=50$, $100$, $250$, $500$ and $1000$. The training sample was formed by generating $50$ observations from each class (except Example~\ref{ex4}) and a test set of size $500$ ($250$ from each class) was used. In Example~\ref{ex4}, the training samples sizes were set to be $50$ and $25$, respectively. This process was repeated $100$ times to compute the average misclassification rates, which are reported in Figure \ref{plot3_a}. For the proposed generalized and block-generalized classifiers, we used $\gamma(t)=1-e^{-t}$ and $\phi(t)=t$.

Observe that in Examples~\ref{ex1}, \ref{ex2}, \ref{ex3}, \ref{ex7} and \ref{ex8}, we have $\bmu_{1d}=\bmu_{2d}=\0_d$ (i.e., $\nu_{12}^2=0$). Furthermore, we have $\sigma_1^2=\sigma_2^2=5/3$ in Example~\ref{ex1} and $\sigma_1^2=\sigma_2^2=0.75$ in Example~\ref{ex7}, while $\sigma_1^2=\sigma_2^2=1$ in Examples~\ref{ex2}, \ref{ex3} and \ref{ex8}. This implies that $\sigma_1^2-\sigma_2^2=0$ for all these five examples. In Example~\ref{ex4}, the moment based quantities $\nu^2_{12}$, $\sigma^2_1$ and $\sigma^2_2$ do not exist as the underlying distributions are Cauchy. On the other hand, Example~\ref{ex5} is a {\it location} problem ($\nu_{12}^2=0.25$ with $\sigma_{1}^2-\sigma_{2}^2=0$), while Example~\ref{ex6} is a {\it scale} problem ($\nu_{12}^2=0$ with $|\sigma_{1}^2-\sigma_{2}^2|=0.5$).
In our earlier analysis of Examples~\ref{ex1}--\ref{ex4}, we assumed the group information $\mathcal{C}$ to be {\it known}. We now analyze all eight examples to validate the fact that the data driven procedure for blocking the variables (developed in Section \ref{cluster}) in combination with the block-generalized classifiers (proposed in Section \ref{Improve}) yield promising performance in high dimensions.

In Examples~\ref{ex1}, \ref{ex4}, \ref{ex5}, \ref{ex6} and \ref{ex7}, the component variables are i.i.d.\ and the populations have differences in their one-dimensional marginals. So, assumptions $(A1)-(A3)$ are satisfied and consequently, the misclassification probabilities of the gSAVG and NN-gMADD classifiers are close to zero (see Figure \ref{plot3_a}). This is not the case for the other three examples. In Examples~\ref{ex2}, \ref{ex3} and \ref{ex8}, the one-dimensional marginals are standard normal for both populations, so assumption $(A3)$ is clearly violated. We observe that both the gSAVG and NN-gMADD classifiers misclassify nearly half of the test points in these examples. On the other hand, assumptions $(A5)-(A7)$ are satisfied for these examples. So, the bgSAVG and NN-bgMADD classifiers classify almost all the test points correctly. 
Blocks of variables were estimated using the method described in Section \ref{cluster}, where we used the absolute value of Pearson's correlation coefficient as the measure of similarity. However, this measure is inappropriate for Example~\ref{ex4} (with Cauchy distributions). So, we have used the minimum regularized covariance determinant (MCD) estimator, which is available through the {\tt R} package {\tt rrcov}.
We observe that the estimated misclassification probabilities of the bgSAVG and NN-bgMADD classifiers are very close to zero in high dimensions (see Figure \ref{plot3_a}), which is consistent with the idea of {\it perfect classification} as $b \to \infty$ (also see Theorem \ref{thm_bgMADD_bgSAVG}).

A question that arises naturally from Figure \ref{plot3_a} is the relative performance of the bgSAVG classifier and the NN-bgMADD classifier for moderate values of $d$. In Section \ref{bgSAVG_and_NN-bgMADD}, we used Examples~\ref{ex2} and \ref{ex4} to motivate this question and investigated this fact theoretically in Theorem \ref{cmpr_AVG_NN}. We now complete this investigation for the other examples. Recall that the relative performance of these two classifiers depends on the ordering of the constants $\tilde \xi_{b}(1,2)$, and $\tilde \tau_{b}(1,2), \tilde \tau_{b}(2,1)$ (see Theorem \ref{cmpr_AVG_NN} and the preceeding discussion).
We have computed the value of these constants in Table \ref{constants_SAVG_NN}. Section~2 of the Supplementary contains more details and related calculations.

We can observe from Figure \ref{plot3_a} that the NN-bgMADD classifier performs better than the bgSAVG classifier in Examples~\ref{ex1}, \ref{ex2}, \ref{ex3} and \ref{ex6} for moderate values of $d$ ($\sim 100-250$). On the contrary, the bgSAVG classifier clearly dominates the NN-bgMADD classifier in Examples~\ref{ex4}, \ref{ex5}, \ref{ex7} and \ref{ex8}. This phenomena is consistent with the ordering of $\tilde \xi_{b}(1,2)$, and $\tilde \tau_{b}(1,2), \tilde \tau_{b}(2,1)$ in Table \ref{constants_SAVG_NN}, except in Examples~\ref{ex5} and \ref{ex7}, where the value of these constants are equal. Interestingly, the bgSAVG classifier performs better than the NN-bgMADD classifier in these two examples. This can be explained by looking closer into the expression of these constants. Recall from Corollary \ref{cor2} that these constants involve the terms $\tilde h_{b}(1,1)$, $\tilde h_{b}(2,2)$ and $\tilde h_{b}(1,2)$. The fact that $\tilde h_{b}(1,2) > \max\{\tilde h_{b}(1,1),\tilde h_{b}(2,2)\}$ (see the values for Examples~\ref{ex5} and \ref{ex7} in Table \ref{constants_SAVG_NN}) justifies the improved performance of the bgSAVG classifier (also see \cite{SBG2020} for related explanations in the context of two sample testing).

\begin{table}[htp]
	\caption{Values of the constants $\tilde \xi_{b}(1,2)$, $\tilde \tau_{b}(1,2)$ and $\tilde \tau_{b}(2,1)$ in Examples~\ref{ex1}--\ref{ex8}. The figure in {\bf bold} indicates the maximum of these three values.} 
	\centering 
	\vspace*{0.1in}
	\begin{tabular}{c c c c c c c} 
		\hline 
		Ex. & $\tilde h_{b}(1,1)$ & $\tilde h_{b}(2,2)$ & $\tilde h_{b}(1,2)$ & $\tilde \xi_{b}(1,2)$ & $\tilde \tau_{b}(1,2)$  & $\tilde \tau_{b}(2,1)$ \rule{0pt}{2.6ex} \\ [0.5ex] 
		\hline 
		1 & 0.6387 & 0.6017 & 0.6230 & 0.0027 & {\bf 0.0185} & {\bf 0.0185} \\
		2 & 0.7909 & 0.6967 & 0.7539 & 0.0101 & 0.0470 & {\bf 0.0472} \\
		\hspace*{0.025in} 3$^*$ & 0.7614 & 0.7091 & 0.7423 & 0.0070 & 0.0260 & {\bf 0.0262} \\
		4 & 0.7440 & 0.6789 & 0.7442 & {\bf 0.0327} & 0.0213 & 0.0222 \\
		5 & 0.5528 & 0.5528 & 0.5583 & {\bf 0.0056} & {\bf 0.0056} & {\bf 0.0056} \\
		6 & 0.5528 & 0.4226 & 0.5000 & 0.0123 & 0.0649 & {\bf 0.0652} \\
		7 & 0.4877 & 0.4877 & 0.5000 & {\bf 0.0123} & {\bf 0.0123} & {\bf 0.0123} \\
		8 & 0.8138 & 0.5903 & 0.7634 & 0.0614 & 0.1111 & {\bf 0.1124} \\
		\hline 
	\end{tabular}
	
	\vspace*{0.05in} $^*$ the block size ($d_i$) was fixed at $5$  
	\label{constants_SAVG_NN}
\end{table}

\subsection{Comparison with popular classifiers} \label{num_popular}

Here, we compare the performance of the proposed classifiers with some well-known classifiers, namely, Support Vector Machines \citep[SVM,][]{vapnik1998statistical}, GLMNET \citep{hastie2009elements}, neural networks \citep[NNET,][]{bishop1995neural} and nearest neighbor classifiers based on the random projection method \citep[NN-RAND,][]{deegalla2006reducing}. We studied numerical performance of these classifiers for $d=1000$ (see Tables 2 and 3 in the Supplementary for other values of $d$). The average misclassification rates along with the corresponding standard errors are reported in Table \ref{simtable}. 
Misclassification rates of both the linear and non-linear SVM are reported. We used the radial basis function (RBF) kernel, i.e., $K_\theta(\bx,\by)=\mathrm{exp}\{-\theta\|\bx-\by\|^2\}$ in non-linear SVM with {$\theta\in\{i/10d; \ 1\leq i\leq 20\}$} and reported the minimum misclassification rate.
For NNET, we used the sigmoid as its activation function. The number of hidden layers were allowed to vary in the set $\{1,3,5,10\}$, and the minimum misclassification rate was reported as NNET. We have used default values for the other parameters that were involved with these classifiers.
The {\tt R} packages {\tt e1071}, {\tt glmnet}, {\tt RSNNS} and {\tt RandPro} were used for SVM, GLMNET, NNET and NN-RAND, respectively. Our classifiers were implemented in {\tt R} too, and the codes are available from \href{https://www.dropbox.com/sh/6wcdxzed8wg2xs3/AABag_Zp9biMOQQwt6XFaiF9a?dl=0}{this link}.
We fix $\phi(t)=t$ for the proposed classifiers.
Untill this point, we have used the choice $\gamma_1(t)=1-e^{-t}$ only. We now introduce two more choices of $\gamma$, namely, $\gamma_2(t)=\log(1+t)$ and $\gamma_3(t)=\sqrt{t}/2$ in this section. For our proposed methods, we report the misclassification rates for all three choices of $\gamma$ in Table \ref{simtable}.

\begin{landscape}
	\begin{table}[htp]
		{\scriptsize
			\centering
			\vspace*{-0.3in}
			\caption{Misclassification rates (stated in the first row) and standard errors (stated in the second row) of different classifiers in Examples~\ref{ex1}--\ref{ex8} for $d=1000$. The figure in {\bf bold} indicates the minimum misclassification rate.}
			\vspace*{0.1in}
			\renewcommand{\arraystretch}{1.2}
			\begin{tabular}{|c|c|c|c|c|c|c@{\hskip4pt}c@{\hskip4pt}c|c@{\hskip4pt}c@{\hskip4pt}c|c@{\hskip4pt}c@{\hskip4pt}c|c@{\hskip4pt}c@{\hskip4pt}c|}
				\hline
				\multicolumn{1}{|l}{Ex.} & \multicolumn{1}{|@{\hskip1.5pt}c@{\hskip1.5pt}}{GLMNET}  &  \multicolumn{1}{|c}{NN}  & \multicolumn{1}{|c}{SVM} & \multicolumn{1}{|c}{SVM}&\multicolumn{1}{|c|}{NNET} & & \multicolumn{1}{@{\hskip2.5pt}c@{\hskip2.5pt}}{gSAVG} & & & \multicolumn{1}{@{\hskip1.5pt}c@{\hskip1.5pt}}{bgSAVG}  & &\multicolumn{3}{@{\hskip1pt}c@{\hskip1pt}|}{NN-gMADD} & \multicolumn{3}{@{\hskip5pt}c@{\hskip2pt}|}{NN-bgMADD} \\
				\multicolumn{1}{|l}{} & \multicolumn{1}{|c}{}  &  \multicolumn{1}{|c}{-RAND}  & \multicolumn{1}{|c}{-LIN} & \multicolumn{1}{|c}{-RBF} &\multicolumn{1}{|c|}{} &\multicolumn{1}{c}{$\gamma_1$} & \multicolumn{1}{c}{$\gamma_2$} & \multicolumn{1}{c}{$\gamma_3$} & \multicolumn{1}{|c}{$\gamma_1$}& \multicolumn{1}{c}{$\gamma_2$}  & \multicolumn{1}{c|}{$\gamma_3$} & \multicolumn{1}{|c}{$\gamma_1$}& \multicolumn{1}{@{\hskip1pt}c@{\hskip1pt}}{$\gamma_2$}  & \multicolumn{1}{c|}{$\gamma_3$}&\multicolumn{1}{|c}{$\gamma_1$}& \multicolumn{1}{c}{$\gamma_2$}  & \multicolumn{1}{c|}{$\gamma_3$}\\
				\hline
				1     & 0.4748  & 0.4972 & 0.4979 & 0.4952 & 0.4919 & 0.1002   & 0.2079 &0.2646 & {0.1167} & 0.2156 & 0.2702  & {\bf 0.0302}  & 0.1321 &0.2451  & 0.0379   & 0.1411 & 0.2457\\ 
				& 0.0177  & 0.0171 & 0.0232 & 0.0203& 0.0240 &   0.0194 & 0.0195 & 0.0208 &  0.0165  & 0.0229 & 0.0230 & 0.0102 & 0.0260 & 0.0314 &  0.0135  & 0.0274 & 0.0374 \\
				\hline
				2     & 0.4745 & 0.4940 & 0.5099 & 0.4540 & 0.5010 & 0.5025 & 0.5029 & 0.5024 &  0.0815  & 0.1243 &0.1461 & 0.4445 & 0.4390 & 0.4384 & 0.0185 & 0.0171  & {\bf 0.0168} \\ 
				& 0.0174 & 0.0150 & 0.0208 & 0.0226 & 0.0253 & 0.0223 & 0.0228 & 0.0224  & 0.0152  & 0.0201 & 0.0208 &  0.0166  & 0.0174 & 0.0173 & 0.0088  & 0.0084 & 0.0084\\
				\hline
				3   & 0.4757 & 0.4558 & 0.5000 & 0.5000 & 0.4997  & 0.4991 & 0.5011 & 0.5018 &  0.0843  & 0.1431 & 0.1532 & 0.4495 & 0.4442 & 0.4443 &  0.0185  & 0.0184 & {\bf 0.0182} \\
				& 0.0182 & 0.0279 & 0.0000 & 0.0000 & 0.0232 & 0.0214 & 0.0230 & 0.0227 & 0.0214  & 0.0260 & 0.0269 & 0.0165 & 0.0152 & 0.0161 &  0.0100  & 0.0105 & 0.0105\\
				\hline
				4     & 0.4173  & 0.4933 & 0.4282  & 0.4995   & 0.3688 & {\bf 0.0000} & {\bf 0.0000} & 0.0017 & {\bf 0.0000}  & {\bf 0.0000} & 0.0022  & {\bf 0.0000}  & 0.0007 & 0.2319 & {\bf 0.0000}  & 0.0009 & 0.2279 \\ 
				& 0.0266 & 0.0245 & 0.0205 & 0.0014 & 0.0236 & 0.0000 & 0.0000 & 0.0018 & 0.0000 & 0.0000  & 0.0021 & 0.0000 & 0.0016  & 0.0341 & 0.0000  & 0.0018 & 0.0313\\
				\hline
				5     & 0.2172  & 0.0336  & 0.0018  & 0.0012  & 0.2748  & 0.0142  & 0.0022 & 0.0018   & 0.0028  & {\bf 0.0007} & {\bf 0.0007}  & 0.1078  & 0.0248 & 0.0202 & 0.0325  & 0.0139 & 0.0134 \\ 
				& 0.0220  & 0.0139 & 0.0020 & 0.0017 & 0.0444 & 0.0055  & 0.0020 & 0.0017 & 0.0028  & 0.0014 & 0.0014 & 0.0261 & 0.0102   & 0.0092  &  0.0173 & 0.0088 &0.0089\\
				\hline
				6     & 0.4533  & 0.5000  & 0.4587 & 0.0000 & 0.4968 & {\bf 0.0000} & {\bf 0.0000} & 0.0003  & {\bf 0.0000} & {\bf 0.0000}  & 0.0003 &  {\bf 0.0000} &  {\bf 0.0000}  & {\bf 0.0000}  &  {\bf 0.0000}  &  {\bf 0.0000} & {\bf 0.0000}  \\ 
				& 0.0158 & 0.0000 & 0.0153 & 0.0000 & 0.0238 &  0.0000  &  0.0000  &  0.0009 &  0.0000  & 0.0003 & 0.0008 &  0.0000  &  0.0000  &  0.0000  &  0.0000  &  0.0000 & 0.0000 \\
				\hline
				7     & 0.4677 & 0.3977 & 0.4974 & 0.4694& 0.4968 & {\bf 0.0000} &  {\bf 0.0000} & 0.0002 & {\bf 0.0000}  & {\bf 0.0000} & 0.0002 & 0.0001   & 0.0034 & 0.0143 & 0.0001  & 0.0034 & 0.0148 \\ 
				& 0.0184 & 0.0245 & 0.0240& 0.0228 & 0.0218 & 0.0000  & 0.0002 & 0.0006 &  0.0000 & 0.0000 & 0.0005 &  0.0005 & 0.0036 & 0.0067 &  0.0004  & 0.0034 & 0.0066\\
				\hline
				8     & 0.4767 & 0.5000 & 0.5010 & 0.2106 & 0.4971  & 0.5001 & 0.4987  & 0.4969 & {\bf 0.0003} & 0.0028 & 0.0033 & 0.4036 & 0.3914 &  0.3883 &  0.0005  & 0.0022 & 0.0024  \\ 
				& 0.0153 & 0.0233 & 0.0208 & 0.0218 & 0.0231& 0.0273  & 0.0328 & 0.0328 & 0.0013 & 0.0050 & 0.0064 & 0.0218 & 0.0245 & 0.0240  & 0.0015 &  0.0042 & 0.0048 \\
				\hline
			\end{tabular}%
			\label{simtable}%
		}
	\end{table}%

	\begin{table}[htp]
	{\scriptsize
		\centering
		\caption{ Misclassification rates (stated in the first row) and standard errors (stated in the second row) of different classifiers in four benchmark data sets. The figure in {\bf bold} indicates the minimum misclassification rate.}
		\vspace*{0.1in}
		\hspace*{-0.25in}
		\renewcommand{\arraystretch}{1.2}
		\begin{tabular}{|c|c|c|c|c|c|c@{\hskip4pt}c@{\hskip4pt}c|c@{\hskip4pt}c@{\hskip4pt}c|c@{\hskip4pt}c@{\hskip4pt}c|c@{\hskip4pt}c@{\hskip4pt}c|}
			\hline
			\multicolumn{1}{|c}{Data} & \multicolumn{1}{|@{\hskip1.5pt}c@{\hskip1.5pt}}{GLMNET}  &  \multicolumn{1}{|c}{NN}  & \multicolumn{1}{|c}{SVM} & \multicolumn{1}{|c}{SVM}&\multicolumn{1}{|c|}{NNET} & & \multicolumn{1}{@{\hskip2.5pt}c@{\hskip2.5pt}}{gSAVG} & & & \multicolumn{1}{@{\hskip1.5pt}c@{\hskip1.5pt}}{bgSAVG}  & &\multicolumn{3}{@{\hskip1pt}c@{\hskip1pt}|}{NN-gMADD} & \multicolumn{3}{@{\hskip5pt}c@{\hskip2pt}|}{NN-bgMADD} \\
			\multicolumn{1}{|l}{} & \multicolumn{1}{|c}{}  &  \multicolumn{1}{|c}{-RAND}  & \multicolumn{1}{|c}{-LIN} & \multicolumn{1}{|c}{-RBF} &\multicolumn{1}{|c|}{} &\multicolumn{1}{c}{$\gamma_1$} & \multicolumn{1}{c}{$\gamma_2$} & \multicolumn{1}{c}{$\gamma_3$} & \multicolumn{1}{|c}{$\gamma_1$}& \multicolumn{1}{c}{$\gamma_2$}  & \multicolumn{1}{c|}{$\gamma_3$} & \multicolumn{1}{|c}{$\gamma_1$}& \multicolumn{1}{@{\hskip1pt}c@{\hskip1pt}}{$\gamma_2$}  & \multicolumn{1}{c|}{$\gamma_3$}&\multicolumn{1}{|c}{$\gamma_1$}& \multicolumn{1}{c}{$\gamma_2$}  & \multicolumn{1}{c|}{$\gamma_3$}\\
			\hline
			CricketX & 0.6553 & 0.5039 & 0.6061 & 0.4154 & 0.6643 & 0.6513 & 0.6500 & 0.6472 & 0.6008 & 0.6215 & 0.6167 & 0.3756 & 0.3907 & 0.3929 & {\bf 0.3326} & 0.3612 & 0.3660 \\
					 & 0.0184 & 0.0228 & 0.0212 & 0.0210 & 0.0263 & 0.0201 & 0.0231 & 0.0220 & 0.0279 & 0.0233 & 0.0250 & 0.0218 & 0.0207 & 0.0211 & 0.0212 & 0.0210 & 0.0222\\
			\hline
			EOGHorizontal & 0.4824 & 0.4141 & 0.4691 & 0.4241 & 0.7280 & 0.7334 & 0.5379 & 0.5028 & 0.7135 & 0.4673 & 0.4684 & 0.8524 & 0.5048 & 0.4998 & 0.8788 & {\bf 0.2938} & 0.3475 \\
			Signal		  & 0.0183 & 0.0241 & 0.0236 & 0.0211 & 0.0458 & 0.0183 & 0.0231 & 0.0201 & 0.0127 & 0.0236 & 0.0236 & 0.0170 & 0.0214 & 0.0254 & 0.0153 & 0.0205 & 0.0181 \\
			\hline
			GSE2685 & 0.2060 & 0.2913 & {\bf 0.1787} & 0.3475 & 0.4013 & 0.5213 & 0.4781 & 0.4763 & 0.4438 & 0.4263 & 0.4175 & 0.3575 & 0.2869 & 0.2381 & 0.4480 & 0.2120 & 0.2873 \\
					& 0.0622 & 0.1091 & 0.0613 & 0.0505 & 0.1081 & 0.1159 & 0.1282 & 0.1252 & 0.1413 & 0.1370 & 0.1442 & 0.0875 & 0.0941 & 0.0887 & 0.1396 & 0.0959 & 0.1104 \\
			\hline
			nutt2003v2  & 0.1993 & 0.4000 & 0.1114 & 0.2100 & 0.4993 & 0.3336 & 0.2150 & 0.1871 & 0.3514 & 0.0871 & {\bf 0.0779} & 0.3686 & 0.1957 & 0.1557 & 0.2593 & 0.1286 & 0.1186 \\
						& 0.1081 & 0.0825 & 0.0769 & 0.1695 & 0.0864 & 0.1264 & 0.1082 & 0.1102 & 0.1039 & 0.0588 & 0.0509 & 0.0951 & 0.0784 & 0.0762 & 0.1229 & 0.0626 & 0.0549 \\
			\hline
		\end{tabular}%
		\label{real4}%
	}
\end{table}%

\end{landscape}

In all the examples (except Example~\ref{ex5}), the competing classifiers GLMNET, NN-RAND, SVM and NNET misclassify almost 50\% of the test sample points. Example~\ref{ex5} involves a {\it location} problem, and all these popular classifiers perform quite well, with SVM having a clear edge over the others, followed closely by NN-RAND. The non-linear classifier SVM-RBF leads to {\it perfect classification} in Example~\ref{ex6} (a {\it scale} problem), and an improved misclassification rate of about 21\% in Example~\ref{ex8} (having differences in their scatter matrices).

To summarize the performance of our classifiers in Table \ref{simtable}, we observe that the proposed bgSAVG and NN-bgMADD classifiers outperform popular classifiers in all examples. 
In Example~\ref{ex1}, the misclassification rates of these classifiers are slightly more than those of the gSAVG and NN-gMADD classifiers, respectively. We have difference in marginal distributions, and it is not necessary to use variable clustering in this example. The same is true for Examples~\ref{ex4} and \ref{ex7} as well, but the misclassification rates of the bgSAVG and NN-bgMADD classifiers are quite similar to those of the gSAVG and NN-gMADD classifiers in these two examples. In fact, the additional error incurred due to estimation of groups is negligible in such cases. Moreover, the block-generalized classifiers improve over the generalized classifiers in Example~\ref{ex5}. These examples clearly show that block-generalized classifiers perform well even when it is not necessary to group the component variables.

\subsection{Comparison among the choices of $\gamma$} \label{num_gamma}

A natural question that arises from Table \ref{simtable} is the choice of $\gamma$ in practice.
We have considered three choices of $\gamma$, namely, $\gamma_1(t)=1-e^{-t}$, $\gamma_2(t)=\log(1+t)$ and $\gamma_3(t)=\sqrt{t}/2$. 
All these functions have non-constant, completely monotone derivatives \citep[see, e.g.,][]{MR0270403, BF10}. These functions are monotonically increasing and there exists a $C>0$ such that these functions satisfy the ordering $\gamma_1(t) < \gamma_2(t) < \gamma_3(t)$ for all $t>C$. The function $\gamma_1$ is clearly bounded, while the other two functions are unbounded. For large $t$, the function $\gamma_2$, although unbounded, stays closer to $\gamma_1$ when compared with the function $\gamma_3$. The main idea behind choosing these functions was to explore the complete spectrum (i.e., bounded, unbounded and in-between), and understand the effectiveness of the choice of the $\gamma$ function in capturing discriminative information from the two class distributions.

We deal with heavy-tailed distributions in Example~\ref{ex4}, and the advantage of using a bounded $\gamma$ is clear here. In this example, generalized classifiers based on $\gamma_1$ outperformed those based on $\gamma_3$. The performance of classifiers based on $\gamma_2$ was quite close to $\gamma_1$. The fact that $\gamma_1$ is a bounded function is necessary here to ensure that assumptions $(A1)$ and $(A2)$ hold. In Example~\ref{ex5} (a {\it location} problem) involving light-tailed distributions, generalized classifiers based on $\gamma_3$ clearly outperform those constructed using $\gamma_1$, while the performance of $\gamma_2$ again lies in-between these two choices. A related phenomena was also observed by \cite{BF10} for location problems, where the authors were interested in non-parametric two sample goodness of fit tests in $\mathbb{R}^d$. Observe that if we fix a classifier (say, bgSAVG) in Table~\ref{simtable}, then either $\gamma_1$ (in Examples~\ref{ex1}--\ref{ex4} and \ref{ex6}--\ref{ex8}) or $\gamma_3$ (in Example~\ref{ex5}) leads to the minimum misclassification rate. From the results of our simulation study in Table \ref{simtable}, there is {\it no clear winner} among these two choices of the $\gamma$ function. So, we recommend using both choices, namely, $\gamma_1$ and $\gamma_3$ to obtain a complete picture of the underlying scenario.

\section{Real Data Analysis} \label{real}

Now, we study the performance of our proposed classifiers on other benchmark data sets from three popular databases, namely, Compcancer database, Microarray database and UCR Time Series Archive (2018). Detailed description of the data sets are available at the respective sources. 
Data sets in the Compcancer and Microarray databases (involving gene expression studies) have a {\it fixed data} with corresponding class labels, while those from the UCR Archive come in two parts, a {\it fixed training set} as well as a {\it fixed test set}. 
For our analysis of the data sets in the Compcancer and Microarray databases, we randomly selected $50$\% of the observations (without replacement) corresponding to each class to form the training set. The rest of the observations were considered as test cases. 
For data sets from the UCR Archive, we combined the available training and test data, and randomly selected $50\%$ of the observations from the combined set to form a new set of training observations, while keeping the proportions of observations from different classes consistent. The other half was considered as the test set. This procedure was repeated $100$ times over different splits of the data set to obtain a stable estimate of the misclassification rate.

Let us start by analyzing the four benchmark data sets mentioned in Section \ref{cluster}.
The numerical results are reported in Table \ref{real4}.
The NN-bgMADD classifier captures information from the group structure and leads to the minimum overall misclassification rate in both {\tt Cricket X} and {\tt EOGHorizontalSignal} data sets.
In the {\tt EOGHorizontalSignal} data, we observed a significant variability in the misclassification rates for different choices of $\gamma$. In fact, $\gamma_1$ (a bounded function) led to a misclassification rate of about 88\%. This deteriorating performance of $\gamma_1$ may be attributed to the fact that this function involves the term $e^{-t}$, which reduces the large differences in componentwise means of the competing classes, while $\gamma_3$ involves the term $\sqrt{t}/2$, and manages to retain this information.
The next two data sets are related to gene expression studies, and the component variables often have differences in their class means.
SVM-LIN yields the lowest misclassification rate, while the NN-bgMADD classifier had the second best performance in the {\tt GSE2685} data set.
The bgSAVG classifier leads to the best performance in the high-dimensional {\tt nutt2003v2} data, followed by the SVM-LIN and NN-bgMADD classifiers. Generally, we observe that block-generalized classifiers perform significantly better than their generalized counterparts in all four data sets. This further establishes the usefulness of such classifiers in real data scenarios.

The Compcancer database has $35$ data sets, while the Microarray database consists of $20$ data sets. We chose data sets with $\min_j n_j \geq 6$, which left us with $31$ data sets from the first database, and $20$ data sets in the second database. 
The {\tt ALLGSE412} data set in the Microarray database has missing values in $29$ observations (out of the $55$ samples) corresponding to $14$ covariates, so we dropped those covariates from all the samples during our analysis. We used $71$ (out of available $85$) data sets from the UCR data base.

\begin{figure}[H]
	\begin{center}
		\captionsetup{justification=centering}
			\includegraphics[width= \linewidth]{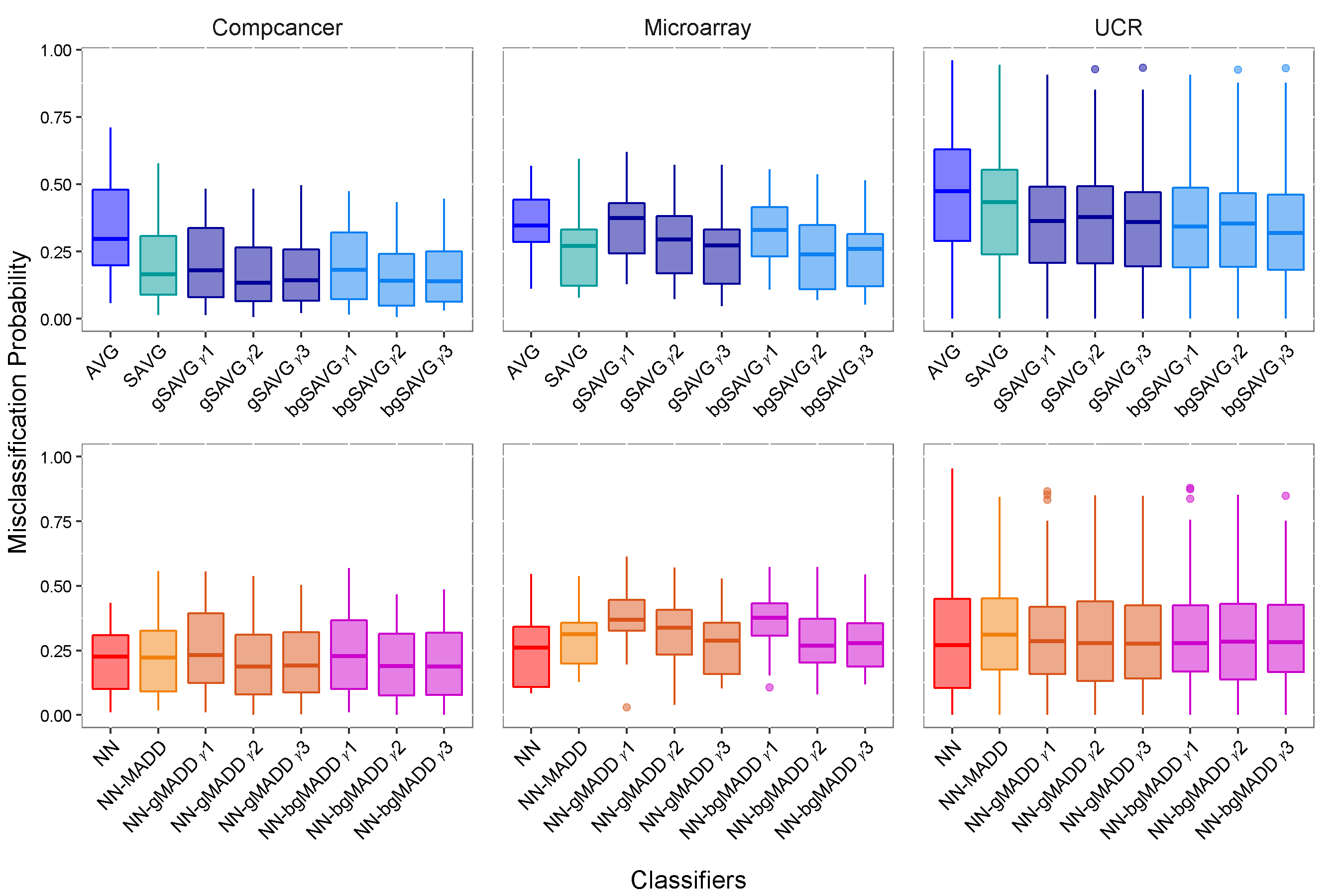}
	\end{center}
	\caption{Boxplot of the estimated misclassification probabilities corresponding to various AVG and NN classifiers in the Compcancer, Microarray and UCR databases.}
	\label{real_data2_AVGNN}
\end{figure}

To begin with, we look at the performance of the generalized and block-generalized classifiers w.r.t.\ their classical counterparts. 
In Figure \ref{real_data2_AVGNN}, we show boxplots of the misclassification rates for the proposed classifiers, separately for the three databases. 
It is clear from these figures that the generalized versions of the AVG classifier yield substantial improvement over the usual classifiers, while the block-generalized classifiers yield further improvement in all three databases.
However, this improvement is not so compelling for the generalized and block-generalized NN classifiers.
Interestingly, {\it simple} classifiers like SAVG and NN yield competitive performance in the first two databases involving gene expression~studies.

\begin{figure}[H]
\vspace*{-0.2in}
	\begin{center}
		\captionsetup{justification=centering}
		\includegraphics[width=0.86\linewidth]{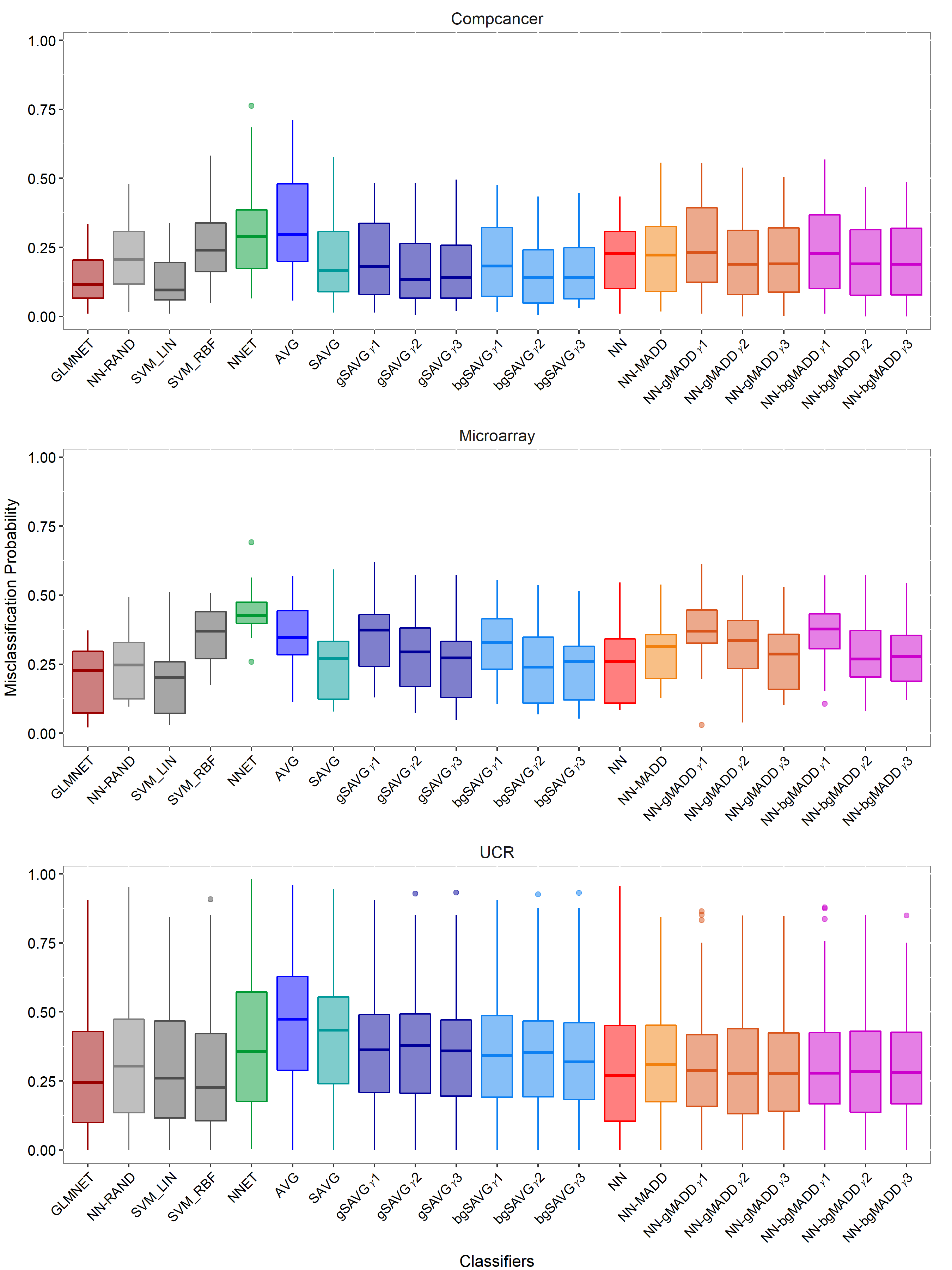}
	\end{center}
	\caption{Boxplot of the estimated misclassification rates corresponding to various classifiers in the Compcancer, Microarray and UCR databases.}
	\label{real_data3}
\end{figure}

Next, we compare the performance of our proposed classifiers with some existing classifiers (namely, SVM, GLMNET, NNET and NN-RAND).
To get an overall picture of their performance in the three databases, 
we summarized the entire information through boxplots in Figure \ref{real_data3} separately for these three databases. 
For each database, we considered a boxplot of misclassification rates for all $22$ classifiers across all data sets in that database. 
Detailed results are available in Section 5 (see Tables 4--11) of the Supplementary.

The Compcancer and Microarray databases have datasets involving gene expressions, which are very high-dimensional ($d \sim 1400-23000$) with low sample sizes ($n \sim 10-100$). Most of these data sets involve $2$ or $3$ class problems. 
Linear SVM performs best in these two databases (see Figure \ref{real_data3}) since the competing classes often have differences in their mean vectors. 
GLMNET (a {\it regularized} linear classifier) induces drastic reduction in the data dimension (the reduced dimension $\sim 1-99$), and takes the second position. These data sets have sparsity in their components, which justifies the good performance of GLMNET. 
However, blocks of variables contain important information (recall panels (c) and (d) of Figure \ref{real_motivation}) and also lead to dimension reduction through the estimated block structure. This helps the bgSAVG classifier to perform quite well too in these two data bases. Generally, the bgSAVG classifier tends to perform better than the NN-bgMADD classifier.

The UCR data archive is quite diverse with $d \sim 24-2700$ and $n \sim 20-700$. The number of classes $J$ varies from $2$ to $52$. 
Again, GLMNET invokes dimension reduction by identifying sparse components, and yields the best performance. Performance of SVM-RBF improves substantially in this database. The NN-bgMADD classifier also performs quite well and secures a competitive position.
Linear classifiers like GLMNET and SVM-LIN perform quite well in data sets with clear differences in their locations, while popular non-linear classifiers like SVM-RBF and NN yield good performance in data sets with difference in scales and/or shapes. In particular, GLMNET and SVM-LIN outperform the non-linear classifiers in the {\tt Coffee} and {\tt Wine} data sets, while SVM-RBF and NN outperform the linear classifiers in the {\tt CinCECGtorso}, {\tt MoteStrain} and {\tt Synthetic Control} data sets.
The NN-bgMADD classifiers seem to have a slight edge over the corresponding bgSAVG classifiers here.
Generally, we observe a large variability in the boxplots for the UCR database because of the presence of data sets with very high as well as low misclassification rates. 
In particular, the {\tt PigAirwayPressure} data with $52$ classes has a misclassification rate of more than $80\%$ across all classifiers, whereas we obtain {\it perfect classification} for these classifiers in the {\tt InsectEPGRgularTrain} data with $3$ classes.

\section{Concluding Remarks} \label{conclude}

In this article, we have studied the HDLSS asymptotic properties of some distance based classifiers. We have analyzed and generalized the popular average distance classifier and the nearest neighbor classifier. On a theoretical note, we have proved that the misclassification probability of the generalized classifiers go to zero (i.e., {\it perfect classification}) in the HDLSS asymptotic regime under very general conditions.
Using a variety of simulated examples and real data sets from three databases, we have amply demonstrated improved performance of the proposed classifiers when compared with a wide variety of popular classifiers.

The idea of clustering of components in Section \ref{Improve} allows us to theoretically explore several possible ways in which $d$ can grow to infinity. In this work, we have considered the case where the block sizes are bounded, while the number of blocks increases with the dimension. One can also keep the number of blocks fixed and allow the size of some (or, all) blocks to grow with $d$. This may lead to concentration of distances within blocks, and the proposed classifiers will then face issues similar to those discussed in \cite{HMN05}. The remaining possibility is to allow both the number of blocks as well as sizes of the blocks to grow to infinity. This, of course, is a complicated setup for theoretical analysis and out of the scope of this article.

Another aspect is handling sparsity in the feature variables. In our theoretical investigations for the generalized classifiers, assumption $(A3)$ corresponds to the case when the number of informative components scales as $d$, but this can be relaxed further (see \cite{SBG2020} for more details). In particular, if the variables are weakly dependent, Theorem \ref{thmgSAVG} can be proved when the number of informative variables scales as $d^{\alpha}$, for some $\alpha > 1/2$. A similar remark holds for assumption $(A7)$ in the context of block-generalized classifiers. In practice, however, one would be interested in capturing the sparse structure in a data dependent way and modify the classifiers accordingly. This is a topic of future research.

\acks{The first and third authors have been partially supported by the DST-SERB grant \sloppy{ECR/2017/000374}. The authors would like to thank the Action Editor for his encouragment, and the three anonymous reviewers for their constructive comments and suggestions that substantially improved the paper.}

\appendix
\section{Proofs and Mathematical Details} \label{AppendixA}

We begin with proofs of the results stated in Section~\ref{Improve}. Proofs of the results in Section \ref{general} are similar, and are in fact special cases (follows by taking $b=d$, equivalently, $d_i=1$ for $1 \leq i \leq d$) of these proofs. Hence, we omit them.

\vspace{0.1in}
\noindent
{\bf Proof of Lemma \ref{secondlemma}} \label{proofofsecondlemma}
Fix $\epsilon>0$. 
Let us define $W_i=\gamma({d_i}^{-1}\|\bU_i-\bV_i\|^2)$ for $1\leq i\leq b$, where $\bU\sim \mathbf{F}_j$ and $\bV \sim \mathbf{F}_{j^\prime}$, $1\leq j,j^\prime\leq J$.
Using Chebyshev's inequality, we observe that 
\begin{align*}
&\Pr\left [\bigg |\frac{1}{b}\sum_{i=1}^b W_i - \frac{1}{b}\sum_{i=1}^b {\rm E}(W_i)\bigg |>\epsilon\right] \leq \frac{1}{\epsilon^2}{\rm E}\left[\frac{1}{b}\sum _{i=1}^b W_i - \frac{1}{b}\sum_{i=1}^b {\rm E}(W_i)\right]^2.
\end{align*}
We are going to show
$${\rm E}\left[\frac{1}{b}\sum _{i=1}^b W_i - \frac{1}{b}\sum_{i=1}^b {\rm E}(W_i)\right]^2 = {\rm Var}\left[\frac{1}{b}\sum _{i=1}^b W_i\right]\to 0 \mbox{ as } b\to\infty.$$
Observe that
\begin{align}\label{converg1}
0&\leq\ {\rm Var}\big [b^{-1}\sum ^b_{i=1}W_i\big ]\\ \nonumber
&=b^{-2}\sum_{i=1}^b {\rm Var}\big [W_i\big ]+2b^{-2}\mathop{\sum\sum}_{1\leq i< i^\prime\leq b} {\rm Cov}\left(W_i,W_{i^\prime}\right)\\ \nonumber
&=b^{-2}\sum_{i=1}^b {\rm Var}\big [W_i\big ]+2b^{-2}\mathop{\sum\sum}_{1\leq i< i^\prime\leq b} {\rm Corr}\left(W_i,W_{i^\prime}\right)\sqrt{{\rm Var}[W_i] {\rm Var}[W_{i^\prime}]}\\ \nonumber
&\leq b^{-2}\sum_{i=1}^b {\rm E}[W^2_i]+2b^{-2}\mathop{\sum\sum}_{1\leq i< i^\prime\leq b} {\rm Corr}\left(W_i,W_{i^\prime}\right)\sqrt{{\rm E}[W^2_i] {\rm E}[W^2_{i^\prime}]}\\ \nonumber 
&\leq b^{-2}\sum_{i=1}^b c_2 + 2 c_2 b^{-2} \mathop{\sum\sum}_{1\leq i< i^\prime\leq b} {\rm Corr}\left(W_i,W_{i^\prime}\right)\hspace{0.5cm}\text{[by assumption $(A5)$]}\\ \nonumber
&\leq c_2 b^{-1} + 2 c_2 b^{-2} \mathop{\sum\sum}_{1\leq i< i^\prime\leq b} {\rm Corr}\left(W_i,W_{i^\prime}\right). \nonumber\\
&=~o(1)\text{   [by assumption $(A6)$].}
\end{align}

\noindent 
Therefore, $\big |b^{-1}\sum_{i=1}^b W_i - b^{-1}\sum_{i=1}^b {\rm E}[W_i]\big | \stackrel{P}{\to} 0\text{ as }b\to\infty$. Since $\phi$ is uniformly continuous, it follows from the definition of uniform continuity that for any $\epsilon_1 > 0$, there exists $\epsilon_2 > 0$ such that
\begin{align*}
& \Pr \big [\big|b^{-1}\sum_{i=1}^b W_i - b^{-1}\sum_{i=1}^b {\rm E}[W_i]\big |\leq\epsilon_2 \big ] \leq \Pr \big [ \big |\phi(b^{-1}\sum_{i=1}^b W_i) - \phi(b^{-1}\sum_{i=1}^b {\rm E}[W_i])\big | \leq \epsilon_1 \big ].
\end{align*} 
Since, $\lim\limits_{b\to\infty}\Pr\big[\big|b^{-1}\sum_{i=1}^b W_i - b^{-1}\sum_{i=1}^b {\rm E}[W_i]\big |\leq\epsilon_2 \big]=1,$
$$\bigg |\phi(b^{-1}\sum_{i=1}^b W_i) - \phi(b^{-1}\sum_{i=1}^b {\rm E}[W_i]) \bigg | \stackrel{P}{\to}0\text{ as }b\to\infty.$$
Hence, $\big|h_b(\bU,\bV)-\tilde{h}_b(j,j^\prime) \big| \stackrel{P}{\to}0$ as $b\to\infty$ for all $1\leq j,j^\prime\leq J.$
\QEDB
\vspace{0.1in}

\noindent
{\bf Proof of Corollary \ref{cor2}} \label{proofofcor2_b}
It follows from Lemma \ref{secondlemma} that for independent random vectors $\bZ \sim \mathbf{F}_j,$ and $\bX,\bX^\prime \stackrel{i.i.d.}{\sim} \mathbf{F}_{j^\prime}$ with $1\leq j,j^\prime\leq J$, we have
$$|h_b(\bZ,\bX)- \tilde{h}_b(j,j^\prime)|\stackrel{P}{\to}0\text{ and }
|h_b(\bX,\bX^\prime)- \tilde{h}_b(j^\prime,j^\prime)|\stackrel{P}{\to}0\text{ as }b\to\infty.$$
This further implies that
\begin{align}
& \bigg |n_{j^\prime}^{-1}\sum\limits_{\bX\in\rchi_j}h_b(\bZ,\bX)- \tilde{h}_b(j,j^\prime) \bigg | \stackrel{P}{\to}0 \text{ and }\nonumber\\ 
& \bigg |\{n_{j^\prime}(n_{j^\prime}-1)\}^{-1}\sum\limits_{\bX,\bX^\prime\in\rchi_j}h_b(\bX,\bX^\prime)- \tilde{h}_b(j^\prime,j^\prime) \bigg | \stackrel{P}{\to}0\text{ as }b\to\infty.\label{hconv}
\end{align}

\begin{enumerate}[(a)]
	\item Recall that for any $1\leq j,j^\prime\leq J,$
	\begin{align*}
	&\xi_{jb}(\bZ)=n_j^{-1}\sum\limits_{\bX\in\rchi_j}h_b(\bZ,\bX)-\{2n_j(n_j-1)\}^{-1}\sum\limits_{\bX,\bX^\prime\in\rchi_j}h_b(\bX,\bX^\prime),\\
	&\xi_{j^\prime b}(\bZ)=n_{j^\prime}^{-1}\sum\limits_{\bX\in\rchi_{j^\prime}}h_b(\bZ,\bX)-\{2n_{j^\prime}(n_{j^\prime}-1)\}^{-1}\sum\limits_{\bX,\bX^\prime\in\rchi_{j^\prime}}h_b(\bX,\bX^\prime),\text{ and }\\
	&\tilde{\xi}_b(j,j^\prime)=\tilde{h}_b(j,j^\prime)- \frac{1}{2}\big (\tilde{h}_b(j^\prime,j^\prime)+\tilde{h}_b(j,j)\big ).
	\end{align*}
	
	Since $\bZ\sim \mathbf{F}_j$, it follows from \eqref{hconv} that
	\begin{equation} \label{3.2.1}
	|\xi_{j^\prime b}(\bZ)-\{\tilde{h}_b(j,j^\prime)- \tilde{h}_b(j^\prime,j^\prime)/2\}\big |\stackrel{P}{\to} 0 \text{ and }
	|\xi_{j b}(\bZ)-\tilde{h}_b(j,j)/2\big |\stackrel{P}{\to}0\text{ as }b\to\infty.  
	\end{equation}
	Consequently,
	\begin{align*}
	&\bigg |\big \{\xi_{j^\prime b}(\bZ)-\xi_{j b}(\bZ)\big \}-\big \{\tilde{h}_b(j,j^\prime)- \frac{1}{2}\big (\tilde{h}_b(j^\prime,j^\prime)+\tilde{h}_b(j,j)\big )\}\bigg |\stackrel{P}{\to}0\text{ as }b\to\infty\\
	\implies&\big |\big \{\xi_{j^\prime b}(\bZ)-\xi_{j b}(\bZ)\big \}-\tilde{\xi}_b(j,j^\prime)\big |\stackrel{P}{\to}0\text{ as }b\to\infty.
	\end{align*}
	
	\item Recall that $\bZ \sim \mathbf{F}_j$ and $\bX \sim \mathbf{F}_{j^\prime}$ with $1\leq j, j^\prime\leq J$, and $\psi_{b}(\bZ,\bX)$ can be expressed as follows: 
	$$\frac{1}{n-1} \Bigg (\sum\limits_{\bX^\prime\in\rchi_{j^\prime}\setminus\{\bX\}} \big|h_b(\bZ,\bX^\prime)-h_b(\bX,\bX^\prime)\big|+\sum\limits_{\bX^\prime\in\rchi\setminus\rchi_{j^\prime}} \big|h_b(\bZ,\bX^\prime)-h_b(\bX,\bX^\prime)\big| \Bigg ).$$ 
	
	\noindent 
	Now, using triangle inequality (repeatedly), we obtain

	\begin{align*}
	0&\leq \big|\psi_{b}(\bZ,\bX)-\tilde{\tau}_b(j,j^\prime)\big|\\
	&=\bigg|\frac{1}{n-1}\bigg \{\sum\limits_{\bX^\prime\in\rchi_{j^\prime}\setminus\{\bX\}} \big|h_b(\bZ,\bX^\prime)-h_b(\bX,\bX^\prime)\big|+\sum\limits_{\bX^\prime\in\rchi\setminus\rchi_{j^\prime}} \big|h_b(\bZ,\bX^\prime)-h_b(\bX,\bX^\prime)\big|\bigg\}\\
	&~~-\bigg\{\frac{n_{j^\prime}-1}{n-1}\mid\tilde{h}_b(j,j^\prime)-\tilde{h}_b(j^\prime,j^\prime)\mid + \sum\limits_{l\neq j^\prime}\frac{n_l}{n-1}\mid\tilde{h}_b(j,l)-\tilde{h}_b(j^\prime,l)\mid\bigg\}\bigg|\\
	&=\bigg|\frac{1}{n-1}\bigg \{\sum\limits_{\bX^{\prime}\in\rchi_{j^\prime}\setminus\{\bX\}} \big|h_b(\bZ,\bX^\prime)-h_b(\bX,\bX^\prime)\big|
	-(n_{j^\prime}-1)\mid\tilde{h}_b(j,j^\prime)-\tilde{h}_b(j^\prime,j^\prime)\mid\\
	&~~+\sum\limits_{\bX^\prime\in\rchi\setminus\rchi_{j^\prime}} \big|h_b(\bZ,\bX^\prime)-h_b(\bX,\bX^\prime)\big| - \sum\limits_{l\neq j^\prime}\frac{n_l}{n-1}\mid\tilde{h}_b(j,l)-\tilde{h}_b(j^\prime,l)\mid\bigg\}\bigg|\\
	&\leq\frac{1}{n-1}\bigg\{\sum\limits_{\bX^\prime\in\rchi_{j^\prime}\setminus\{\bX\}}\bigg|\big|h_b(\bZ,\bX^\prime)-h_b(\bX,\bX^\prime)\big|-\mid\tilde{h}_b(j,j^\prime)-\tilde{h}_b(j^\prime,j^\prime)\mid\bigg|\\
	&~~~~+\sum\limits_{l\neq j^\prime}\sum\limits_{\bX^\prime\in\rchi_l}\bigg|\big|h_b(\bZ,\bX^\prime)-h_b(\bX,\bX^\prime)\big|-\mid\tilde{h}_b(j,l)-\tilde{h}_b(j^\prime,l)\mid\bigg|\bigg\}\\
	&\leq\frac{1}{n-1}\bigg\{\sum\limits_{\bX^\prime\in\rchi_{j^\prime}\setminus\{\bX\}}\big|h_b(\bZ,\bX^\prime)-\tilde{h}_b(j,j^\prime)\big|+ \sum\limits_{\bX^\prime\in\rchi_{j^\prime}\setminus\{\bX\}}\big|h_b(\bX,\bX^\prime)-\tilde{h}_b(j^\prime,j^\prime)\big|\\
	&~~+\sum\limits_{l\neq j^\prime}\sum\limits_{\bX^\prime\in\rchi_l}\big|h_b(\bZ,\bX^\prime)-\tilde{h}_b(j,l)\big|+\sum\limits_{l\neq j^\prime}\sum\limits_{\bX^\prime\in\rchi_l}\big|h_b(\bX,\bX^\prime)-\tilde{h}_b(j^\prime,l)\big|\bigg\}.
	\end{align*}
	It follows from Lemma \ref{secondlemma} that each of the summands converge to $0$ in probability as $b\to\infty.$ Therefore, for a fixed sample size $n$, $\big|\psi_{b}(\bZ,\bX)-\tilde{\tau}_b(j,j^\prime)\big|\stackrel{P}{\to}0\mbox{ as }b\to\infty$ for all $1\leq j, j^\prime\leq J$.
	
	Let us assume that $j\neq j^\prime$. We have $\tau_{jb}(\bZ)=\min_{\bX\in\rchi_j}\psi_{b}(\bZ,\bX),$ and $\tau_{j^\prime b}(\bZ)=\min_{\bX\in\rchi_{j^\prime}} \psi_{b}(\bZ,\bX).$ Since $\bZ \sim \mathbf{F}_{j}$, we get
	$$\big |\tau_{j^\prime b}(\bZ) - \tilde{\tau}_b(j,j^\prime)\big|\stackrel{P}{\to}0 \text{ and}
	\big |\tau_{jb}(\bZ) -\tilde{\tau}_b(j,j)\big|\stackrel{P}{\to}0\mbox{ as }b\to\infty.$$
	Since $\tilde{\tau}_b(j,j)=0$, it follows that
	\begin{align*}
	\big |\{\tau_{j^\prime b}(\bZ)-\tau_{jb}(\bZ)\} -\tilde{\tau}_b(j,j^\prime)\big|\stackrel{P}{\to}0\mbox{ as }b\to\infty.
	\end{align*}
\QEDB
\end{enumerate}

\noindent
{\bf Proof of Lemma \ref{fourthlemma}} \label{proofoflemma3.5}
Suppose that $\bX_1,\bX_2$ are i.i.d. copies of $\bX \sim \mathbf{F}_j,$ and $\bX_3,\bX_4$ are i.i.d. copies of $\bX^\prime \sim \mathbf{F}_{j^\prime}$ for $1\leq j\neq j^\prime\leq J$. Let us denote $\tilde{h}_b(j,j) = \phi(A_{1b}),~\tilde{h}_b(j^\prime,j^\prime) = \phi(A_{2b})$ and $\tilde{h}_b(j,j^\prime) = \phi(A_{3b})$, where $A_{1b} = b^{-1}\sum_{i=1}^b {\rm E}\big [\gamma(d_i^{-1}\|{\bf X}_{1i} - {\bf X}_{2i}\|^2)\big ]$, $A_{2b} = b^{-1}\sum_{i=1}^b {\rm E}\big [\gamma(d_i^{-1}\|{\bf X}_{3i} - {\bf X}_{4i}\|^2)\big ]$ and
$A_{3b} = b^{-1}\sum_{i=1}^b {\rm E}\big [\gamma(d_i^{-1}\|{\bf X}_{1i} - {\bf X}_{3i}\|^2)\big ]$.

\begin{enumerate}[(a)]
\item  
For $1\leq i\leq b$ and $1\leq j \neq j^\prime\leq J$, we have 
$$e(\mathbf{F}_{j,i},\mathbf{F}_{j^\prime,i}) = {\rm E}[\gamma(d_i^{-1}\|\bX_{1i}-\bX_{3i}\|^2)]-\frac{1}{2} \bigg \{{\rm E}[\gamma(d_i^{-1}\|\bX_{1i}-\bX_{2i}\|^2)]+{\rm E}[\gamma(d_i^{-1}\|\bX_{3i}-\bX_{4i}\|^2)] \bigg\}$$ 
is the energy distance between the distributions $\mathbf{F}_{j,i}$ and $\mathbf{F}_{j^\prime, i}$. 
\cite{BF10} showed that the energy distance between two distributions is always non-negative, i.e., $e(\mathbf{F}_{j,i}, \mathbf{F}_{j^\prime,i})\geq 0$, for all $1\leq i\leq b$ and $1\leq j \neq j^\prime\leq J$.
Therefore, 
$${\rm E}[\gamma(d_i^{-1}\|\bX_{1i}-\bX_{3i}\|^2)] \geq \frac{1}{2} \bigg \{{\rm E}[\gamma(d_i^{-1}\|\bX_{1i}-\bX_{2i}\|^2)]+{\rm E}[\gamma(d_i^{-1}\|\bX_{3i}-\bX_{4i}\|^2)] \bigg \},~\forall 1\leq i\leq b.$$
This implies that $A_{3b} \ge \frac{1}{2} (A_{1b} + A_{2b})$. Since $\phi$ is increasing and concave, we have $\phi(A_{3b}) \ge \phi\big(\frac{1}{2}A_{1b} + \frac{1}{2}A_{2b}\big) \ge \frac{1}{2}\phi(A_{1b}) + \frac{1}{2}\phi(A_{2b})$. This further implies that $\tilde{\xi}_b(j,j^\prime) = \tilde{h}_b(j,j^\prime) - \frac{1}{2} \big \{\tilde{h}_b(j,j) + \tilde{h}_b(j^\prime,j^\prime) \big \} \geq 0$.

\cite{BF10} also showed that $e(\mathbf{F}_{j,i}, \mathbf{F}_{j^\prime,i})= 0$ if and only if $\mathbf{F}_{j,i}=\mathbf{F}_{j^\prime,i}$, and we have $\tilde{\xi}_b(j,j^\prime)=0$. 
So, we have $\phi(A_{3b}) = \frac{1}{2}\phi(A_{1b}) + \frac{1}{2}\phi(A_{2b})$. Since $\phi$ is concave and increasing, it is straightforward to check that $\frac{1}{2}A_{1b} + \frac{1}{2}A_{2b} \geq A_{3b}$. But, we already know that $A_{3b}\geq \frac{1}{2}A_{1b} + \frac{1}{2}A_{2b}$ and hence, the equality follows. 

This further implies that $\frac{1}{b}\sum_{i=1}^b e(\mathbf{F}_{j,i}, \mathbf{F}_{j^\prime,i})=0\text{ for all }1\leq j\neq j^\prime\leq J$, i.e., $e(\mathbf{F}_{j,i}, \mathbf{F}_{j^\prime,i})$ $= 0\text{ for all }1\leq i\leq b\text{ and }1\leq j\neq j^\prime\leq J$. Clearly, $\mathbf{F}_{j,i}=\mathbf{F}_{j^\prime,i}\text{ for all }1\leq i\leq b\text{ and }1\leq j\neq j^\prime\leq J$ now follows.

Let us assume that $\mathbf{F}_{j,i}=\mathbf{F}_{j^\prime,i}$ for all $1\leq i\leq b$ and $1\leq j\neq j^\prime\leq J$. 
Therefore, we get 
$${\rm E} \big [\gamma \big ({d_i}^{-1}\|\bX_{1i}-\bX_{2i}\|^2 \big ) \big ]={\rm E} \big [\gamma \big ({d_i}^{-1}\|\bX_{1i}-\bX_{3i}\|^2 \big ) \big ] = {\rm E} \big [\gamma \big ({d_i}^{-1}\|\bX_{3i}-\bX_{4i}\|^2 \big ) \big ]$$
which implies that $A_{1b} = A_{2b} = A_{3b}$.
As a consequence, we obtain $\tilde{h}_b(j,j)=\tilde{h}_b(j,j^\prime)=\tilde{h}_b(j^\prime,j^\prime)$, and hence $\tilde{\xi}_b(j,j^\prime)=0$ for $1\leq j\neq j^\prime\leq J$.

\item Recall that for $1\leq j\neq j^\prime\leq J$, we have
$$\tilde{\tau}_b(j,j^\prime)=\frac{n_{j^\prime}-1}{n-1}\mid\tilde{h}_b(j,j^\prime)-\tilde{h}_b(j^\prime,j^\prime)\mid + 
{\sum \limits_{l\neq j^\prime}} \frac{n_l}{n-1}\mid\tilde{h}_b(j,l)-\tilde{h}_b(j^\prime,l)\mid\geq 0.$$ 
If $\tilde{\tau}_b(j,j^\prime)=0$, then $\tilde{h}_b(j,l)=\tilde{h}_b(j^\prime,l)\text{ for all }1\leq l\leq J$. So, we get $\tilde{h}_b(j,j)=\tilde{h}_b(j,j^\prime)=\tilde{h}_b(j^\prime,j^\prime)~[\because \tilde{h}_b(j,j^\prime)=\tilde{h}_b(j^\prime,j)]$. This further implies $\phi (A_{1b} )=\phi(A_{2b}) = \phi(A_{3b})$, and since $\phi$ is one-to-one, we get $A_{1b} =A_{2b} = A_{3b}$. So, we have $\tilde{\xi}_b(j,j^\prime)=A_{3b}-\frac{1}{2}\{A_{1b}+A_{2b}\}=0$.
This implies $\mathbf{F}_{j,i}=\mathbf{F}_{j^\prime,i}\text{ for all }1\leq i\leq b$.

Let us now assume that $\mathbf{F}_{j,i}=\mathbf{F}_{j^\prime,i}$ for all $1\leq i\leq b.$ Consequently, for $\bX^\prime \sim \mathbf{F}_l$ with $1\leq l\leq J \text{ and }1\leq i\leq b$, we get the following
\begin{align*}
&{\rm E}\big [\gamma\big ({d_i}^{-1}\|\bX_{1i}-\bX^\prime_{i}\|^2 \big ) \big ]={\rm E} \big [\gamma \big ( {d_i}^{-1}\|\bX_{3i}-\bX^\prime_{i}\|^2 \big ) \big ]~\\
\implies&\phi \bigg (b^{-1}\sum\limits_{i=1}^b {\rm E} \big [\gamma \big ({d_i}^{-1}\|\bX_{1i}-\bX^\prime_{i}\|^2 \big ) \big ]\bigg )=\phi\bigg (b^{-1}\sum\limits_{i=1}^b {\rm E} \big [\gamma \big ({d_i}^{-1}\|\bX_{3i}-\bX^\prime_{i}\|^2 \big ) \big ] \bigg )\\
\implies&\tilde{\tau}_b(j,j^\prime)=0.\nonumber
\end{align*}
This completes the proof. \QEDB
\end{enumerate}
\vspace{0.01in}

\noindent
Recall that assumption $(A7)$ implies $\liminf_{b\to\infty} \tilde{\tau}_b(j,j^\prime)>0$ for any $1 \leq j \neq j^\prime \leq J$. We now state and prove this fact below. 
\begin{lemma} \label{xi_implies_phi}
If $\liminf_{b\to\infty} \tilde{\xi}_b^{\phi,\gamma}(j,j^\prime)>0$, then we have $\liminf_{b\to\infty} \tilde{\tau}_b^{\phi,\gamma}(j,j^\prime)>0$ for any $1 \leq j \neq j^\prime \leq J$. 
\end{lemma}

\noindent
{\bf Proof of Lemma \ref{xi_implies_phi}} \label{proofofxi_implies_phi}
Recall that 
\begin{align*}
&\tilde{\xi}_b(j, j^\prime)=\tilde{h}_b(j, j^\prime)-\frac{1}{2}\big [\tilde{h}_b(j, j)+\tilde{h}_b(j^\prime, j^\prime)\big ]\text{, and }\\
&\tilde{\tau}_b(j, j^\prime) = {\sum\limits_{l\neq j^\prime}} \bigg \{ \frac{n_l}{n-1}|\tilde{h}_b(j,l)-\tilde{h}_b(j^\prime,l)| \bigg \}+\frac{n_{j^\prime}-1}{n-1}|\tilde{h}_b(j, j^\prime)-\tilde{h}_b(j^\prime, j^\prime)|.
\end{align*}
Since
\begin{align*}
\tilde{\xi}_b(j, j^\prime)&=\tilde{h}_b(j, j^\prime)-\frac{1}{2}\big [\tilde{h}_b(j, j)+\tilde{h}_b(j^\prime, j^\prime)\big ]\\
&=\frac{1}{2}\big [\tilde{h}_b(j, j^\prime)-\tilde{h}_b(j, j)\big ] + \frac{1}{2}\big [\tilde{h}_b(j, j^\prime)-\tilde{h}_b(j^\prime, j^\prime)\big ]\\
&\leq\frac{1}{2}\big |\tilde{h}_b(j, j^\prime)-\tilde{h}_b(j, j)\big | + \frac{1}{2}\big |\tilde{h}_b(j, j^\prime)-\tilde{h}_b(j^\prime, j^\prime)\big |,
\end{align*}
it follows that $$\liminf_{b\to\infty}\tilde{\xi}_b(j, j^\prime)>0
\implies\liminf_{b\to\infty}\bigg ( \frac{1}{2}\big |\tilde{h}_b(j, j^\prime)-\tilde{h}_b(j, j)\big | + \frac{1}{2}\big |\tilde{h}_b(j, j^\prime)-\tilde{h}_b(j^\prime, j^\prime)\big |\bigg )>0.$$
Now, let us assume that 
$$\liminf_{b\to\infty} \bigg ( \frac{1}{2}\big |\tilde{h}_b(j, j^\prime)-\tilde{h}_b(j, j)\big | + \frac{1}{2}\big |\tilde{h}_b(j, j^\prime)-\tilde{h}_b(j^\prime, j^\prime)\big | \bigg )=c,$$
for some $c>0.$
This means that for any $\epsilon >0$, there exists a $b(\epsilon)$ such that for all $b \geq b(\epsilon)$, we have
\begin{align*}
&\frac{1}{2}\big |\tilde{h}_b(j, j^\prime)-\tilde{h}_b(j, j)\big | + \frac{1}{2}\big |\tilde{h}_b(j, j^\prime)-\tilde{h}_b(j^\prime, j^\prime)\big |>c-\epsilon\\
\implies &\frac{1}{2}\big |\tilde{h}_b(j, j^\prime)-\tilde{h}_b(j, j)\big | >\frac{c-\epsilon}{2},\text{ or } \frac{1}{2}\big |\tilde{h}_b(j, j^\prime)-\tilde{h}_b(j^\prime, j^\prime)\big |>\frac{c-\epsilon}{2}\\
\implies&\frac{n_j}{n-1}\big |\tilde{h}_b(j, j^\prime)-\tilde{h}_b(j, j)\big | +\frac{n_{j^\prime}-1}{n-1}\big |\tilde{h}_b(j, j^\prime)-\tilde{h}_b(j^\prime, j^\prime)\big |\\ 
&> \min \bigg \{\frac{n_j(c-\epsilon)}{n-1},\frac{(n_{j^\prime}-1)(c-\epsilon)}{n-1} \bigg \}\\
\implies&\tilde{\tau}_b(j, j^\prime)>\min \bigg \{\frac{n_j(c-\epsilon)}{n-1},\frac{(n_{j^\prime}-1)(c-\epsilon)}{n-1} \bigg \}.
\end{align*}
Since $\epsilon$ is chosen arbitrarily, we obtain the following
$$\liminf_{b\to\infty}\tilde{\tau}_b(j, j^\prime) > c \min \bigg \{\frac{n_j}{n-1},\frac{n_{j^\prime}-1}{n-1} \bigg \}>0.$$
Similarly, it can be shown that 
$$\liminf_{b\to\infty} \tilde{\tau}_b(j^\prime,j) > c \min \bigg \{\frac{n_j-1}{n-1},\frac{n_{j^\prime}}{n-1} \bigg \}>0.$$
This completes the proof. \QEDB
\vspace{0.1in}

\newpage
\noindent
{\bf Proof of Theorem \ref{thm_bgMADD_bgSAVG}} \label{proofofthmbgMADD}
\begin{enumerate}[(a)]
	\item The misclassification probability of the bgSAVG classifier is defined as 
	$$ \Delta_{\rm bgSAVG}=\Pr[\delta_{\rm bgSAVG}({\bf Z}) \neq Y],$$
	where $Y$ denotes the true label of ${\bf Z}$. We will prove that $\Delta_{\rm bgSAVG} \to 0$ as $b\to\infty$. Now, note that
	\begin{align} \label{main_proof}
	0\leq &\lim_{b\to\infty}\Pr[\delta_{\rm bgSAVG}({\bf Z})\neq Y]\nonumber\\ 
	=& \lim_{b\to\infty}\sum\limits_{j=1}^J \Pr[\delta_{\rm bgSAVG}({\bf Z})\neq j,\bZ \sim \mathbf{F}_j]\nonumber\\
	=&\sum\limits_{j=1}^J \pi_j \lim_{b\to\infty}\Pr[\delta_{\rm bgSAVG}({\bf Z})\neq j\mid\bZ \sim \mathbf{F}_j]\nonumber\\
	=& \sum\limits_{j=1}^J \pi_j \lim_{b\to\infty} \Pr[\xi_{j b}({\bf Z}) -\xi_{j^\prime b}({\bf Z})>0~\text{for some } j^\prime \neq j, 1\leq j^\prime \leq J\mid\bZ\sim \mathbf{F}_j]\nonumber \\
	\leq & \sum\limits_{j=1}^J \pi_j \lim_{b\to\infty} \sum\limits_{1\leq j\neq j^\prime\leq J}\Pr[\xi_{j b}({\bf Z}) -\xi_{j^\prime b}({\bf Z})>0\mid\bZ\sim \mathbf{F}_j]\nonumber \\
	=& \sum\limits_{j=1}^J \pi_j \sum\limits_{1\leq j\neq j^\prime\leq J}\lim_{b\to\infty}\Pr[\xi_{j b}({\bf Z}) -\xi_{j^\prime b}({\bf Z})>0\mid\bZ\sim \mathbf{F}_j].
	\end{align}
	
	For any $\theta>0$ and $\epsilon>0$, there exists a $B_1$ such that for all $b\geq B_1$, we have
	\begin{align*}
	&\Pr[|\xi_{j^\prime b}({\bf Z}) - \xi_{jb}({\bf Z})-\tilde{\xi}_b(j,j^\prime)|<\theta\mid\bZ\sim \mathbf{F}_j]>1-\epsilon\ [\text{see Corollary \ref{cor2}(a)}]\\
	\implies&\Pr[\xi_{j^\prime b}({\bf Z}) - \xi_{jb}({\bf Z})-\tilde{\xi}_b(j,j^\prime)>-\theta\mid\bZ\sim \mathbf{F}_j]>1-\epsilon\\
	\implies&\Pr[\xi_{j^\prime b}({\bf Z}) - \xi_{jb}({\bf Z})>-\theta+\tilde{\xi}_b(j,j^\prime)\mid\bZ\sim \mathbf{F}_j]>1-\epsilon.
	\end{align*}
	Let $\liminf_b{\tilde{\xi}_b(j,j^\prime)}$ be denoted by $\tilde{\xi}(j,j^\prime)$. For any $\theta^\prime>0$, there exists a $B^\prime$ such that $\tilde{\xi}_b(j,j^\prime)>\xi(j,j^\prime)-\theta^\prime$ for all $b\geq B^\prime$. Therefore,
	\begin{align}
	&\Pr[\xi_{j^\prime b}({\bf Z}) - \xi_{jb}({\bf Z})>-\theta+\tilde{\xi}_b(j,j^\prime)\mid\bZ\sim \mathbf{F}_j]\nonumber\\
	&\leq \Pr[\xi_{j^\prime b}({\bf Z}) - \xi_{jb}({\bf Z})>-\theta-\theta^\prime+\tilde{\xi}(j,j^\prime)\mid\bZ\sim \mathbf{F}_j]\text{ for all }b\geq B^\prime \nonumber\\
	\implies&\Pr[\xi_{j^\prime b}({\bf Z}) - \xi_{jb}({\bf Z})>-\theta-\theta^\prime+\tilde{\xi}(j,j^\prime)\mid\bZ\sim \mathbf{F}_j]>1-\epsilon\text{ for all }b\geq \max \{B^\prime,B_1\}.\label{theta_{a_1}rbit1}
	\end{align}
	
	Since $\theta,\theta^\prime$ are arbitrary, it can be concluded from equation \eqref{theta_{a_1}rbit1} that
	\begin{align}
	&\lim\limits_{b\to\infty}\Pr[\xi_{j^\prime b}({\bf Z}) - \xi_{jb}({\bf Z})\geq\tilde{\xi}(j,j^\prime)\mid\bZ\sim \mathbf{F}_j]=1 \nonumber\\
	\implies&\lim\limits_{b\to\infty}\Pr[\xi_{j^\prime b}({\bf Z}) - \xi_{jb}({\bf Z})>0\mid\bZ\sim \mathbf{F}_j]=1~ [\because \tilde{\xi}(j,j^\prime)>0] \nonumber\\
	\implies&\lim\limits_{b\to\infty}\Pr[\xi_{j b}({\bf Z}) - \xi_{j^\prime b}({\bf Z})>0\mid\bZ\sim \mathbf{F}_j]=0.\label{lab1}
	\end{align}
	Now, it follows from equations \eqref{main_proof} and \eqref{lab1} that 
	$$\lim\limits_{b\to\infty}\Pr[\delta_{\rm bgSAVG}(\bZ)\neq Y] =\sum\limits_{j=1}^J \pi_j\cdot 0 =0.$$
	
	\item Proof for the misclassification probability of the NN-bgMADD classifier is similar, and follows along the lines of the proof of part (a). 
	Please check Section 1 of the Supplementary for a proof. \QEDB 
\end{enumerate}
\vspace{0.1in}

\noindent
{\bf Proof of Theorem \ref{cmpr_AVG_NN}}
Suppose $0\leq s_i, t_i \leq 1,$ for $1\leq i\leq K.$ Then
\begin{align}
\prod\limits_{i=1}^K(s_i+t_i) &= \sum\limits_{S\subseteq\{1,\ldots, K\}}\prod\limits_{i\in S}s_i \prod\limits_{i\in \{1,\ldots, K\}\setminus S} t_i\nonumber\\
&=\prod\limits_{i=1}^K s_i + \sum\limits_{S\subset\{1,\ldots, K\}}\prod\limits_{i\in S}s_i \prod\limits_{i\in \{1,\ldots, K\}\setminus S} t_i\nonumber\\
&\leq\prod\limits_{i=1}^K s_i + \sum\limits_{S\subset\{1,\ldots, K\}}\prod\limits_{i\in \{1,\ldots, K\}\setminus S} t_i\nonumber\\
&\leq\prod\limits_{i=1}^K s_i + \sum\limits_{i\in\{1,\ldots, K\}}C_K~ t_i,\label{identity}
\end{align}
for some appropriate constant $C_K>0$.

\noindent
Recall that
$\Delta_{\rm bgSAVG} = 1 - \Pr[\delta_{\rm bgSAVG}(\bZ)= Y]$ and $\Delta_{\rm NN-bgMADD} = 1 - \Pr[\delta_{\rm NN-bgMADD}(\bZ)=Y]$. Here,
$$\Pr[\delta_{\rm bgSAVG}(\bZ)=Y]
=\sum_{j=1}^J\pi_j \Pr[\xi_{j^\prime b}(\bZ)-\xi_{j b}(\bZ)>0~\forall j^\prime\neq j, 1\leq j^\prime \leq J|\bZ\sim \mathbf{F}_j],\text{ and }$$
$$\Pr[\delta_{\rm NN-bgMADD}(\bZ)=Y]
=\sum_{j=1}^J \pi_j \Pr[\tau_{j^\prime b}(\bZ)-\tau_{jb}(\bZ)>0~\forall j^\prime\neq j, 1\leq j^\prime \leq J|\bZ \sim \mathbf{F}_j].$$

\noindent 
It is to be noted that given $\bZ$ and $\rchi_j$ (training data of the $j$-th class), $\tau_{k b}(\bZ)-\tau_{jb}(\bZ)$ and $\tau_{l b}(\bZ)-\tau_{jb}(\bZ)$ are independently distributed for all $1 \leq k \neq l \leq J, k, l \neq j.$  Therefore, for any $1\leq  j\leq J,$ we can write the following
\begin{align}\label{thm3.5}
&\Pr[\tau_{j^\prime b}(\bZ)-\tau_{jb}(\bZ)>0~\forall j^\prime\neq j, 1\leq j^\prime \leq J|\bZ \sim \mathbf{F}_j]\nonumber\\
=&~{\rm E} \big \{\Pr[\tau_{j^\prime b}(\bZ)-\tau_{jb}(\bZ)>0~\forall j^\prime\neq j, 1\leq j^\prime \leq J|\bZ\sim \mathbf{F}_j, \rchi_j]\big \}\nonumber\\
=&~{\rm E}\bigg \{\prod\limits_{1\leq j^\prime\neq j\leq J}
\Pr[\tau_{j^\prime b}(\bZ)-\tau_{jb}(\bZ)>0|\bZ \sim \mathbf{F}_j,\rchi_j]\bigg \} \nonumber\\
=&~{\rm E} \bigg \{\prod\limits_{1\leq j^\prime\neq j\leq J}\big (\Pr[\tau_{j^\prime b}(\bZ)-\tau_{jb}(\bZ)>0,~\xi_{j^\prime b}(\bZ)-\xi_{j b}(\bZ)>0|\bZ \sim \mathbf{F}_j,\rchi_j]\nonumber\\
&\hspace{0.5cm}+ \Pr[\tau_{j^\prime b}(\bZ)-\tau_{jb}(\bZ)>0, \xi_{j^\prime b}(\bZ)-\xi_{j b}(\bZ)<0|\bZ\sim \mathbf{F}_j, \rchi_j]\big )\bigg \}\nonumber\\
\leq &~{\rm E}\bigg \{\prod\limits_{1\leq j^\prime\neq j\leq J} \big ( \Pr[\xi_{j^\prime b}(\bZ)-\xi_{j b}(\bZ)>0|\bZ\sim \mathbf{F}_j,\rchi_j]\nonumber \\[5pt]
&\hspace{0.5cm}+ \Pr[\{\xi_{j^\prime b}(\bZ)-\xi_{j b}(\bZ)\}- \{\tau_{j^\prime b}(\bZ)-\tau_{jb}(\bZ)\}<0|\bZ\sim \mathbf{F}_j,\rchi_j]\big )\bigg \} \nonumber \\[10pt]
\leq &~{\rm E}\bigg \{\prod\limits_{1\leq j^\prime\neq j\leq J}\big (\Pr[\xi_{j^\prime b}(\bZ)-\xi_{j b}(\bZ)>0|\bZ\sim \mathbf{F}_j,\rchi_j]\big )\nonumber\\
&\hspace{0.5cm}+\sum\limits_{1\leq j^\prime\neq j\leq J} C_J\cdot \Pr[\{\xi_{j^\prime b}(\bZ)-\xi_{j b}(\bZ)\}- \{\tau_{j^\prime b}(\bZ)-\tau_{jb}(\bZ)\}<0|\bZ\sim \mathbf{F}_j,\rchi_j]\bigg \}\text{ [using }\eqref{identity}]\nonumber\\
= &~{\rm E}\bigg \{\prod\limits_{1\leq j^\prime\neq j\leq J}\big (\Pr[\xi_{j^\prime b}(\bZ)-\xi_{j b}(\bZ)>0|\bZ\sim \mathbf{F}_j,\rchi_j]\big )\bigg \}\nonumber\\
&\hspace{0.5cm}+~{\rm E}\bigg \{\sum\limits_{1\leq j^\prime\neq j\leq J} C_J\cdot \Pr[\{\xi_{j^\prime b}(\bZ)-\xi_{j b}(\bZ)\}- \{\tau_{j^\prime b}(\bZ)-\tau_{jb}(\bZ)\}<0|\bZ\sim \mathbf{F}_j,\rchi_j]\bigg \}\nonumber\\
= &~{\rm E}\bigg \{\prod\limits_{1\leq j^\prime\neq j\leq J}\big (\Pr[\xi_{j^\prime b}(\bZ)-\xi_{j b}(\bZ)>0|\bZ\sim \mathbf{F}_j,\rchi_j]\big )\bigg \}\nonumber\\
&\hspace{0.5cm}+\sum\limits_{1\leq j^\prime\neq j\leq J} C_J\cdot \Pr[\{\xi_{j^\prime b}(\bZ)-\xi_{j b}(\bZ)\}- \{\tau_{j^\prime b}(\bZ)-\tau_{jb}(\bZ)\}<0|\bZ\sim \mathbf{F}_j].
\end{align}

For $\bZ \sim \mathbf{F}_j$ and $1\leq j^\prime\neq j\leq J$, using Corollary \ref{cor2}, we have
$$\big|\big\{\xi_{j^\prime b}(\bZ)-\xi_{j b}(\bZ)\big\}-\tilde{\xi}_b(j,j^\prime)\big|\stackrel{\Pr}{\to}0 \mbox{ and }
\big|\{\tau_{j^\prime b}(\bZ)-\tau_{jb}(\bZ)\}-\tilde{\tau}_b(j,j^\prime)\big|\stackrel{\Pr}{\to}0\text{ as }b\to\infty.$$
This now implies that
$$\big|\big\{\xi_{j^\prime b}(\bZ)-\xi_{j b}(\bZ)\big\}-\{\tau_{j^\prime b}(\bZ)-\tau_{jb}(\bZ)\} - \{\tilde{\xi}_b(j,j^\prime)-\tilde{\tau}_b(j,j^\prime)\}\big|\stackrel{\Pr}{\to}0\text{ as }b\to\infty.$$
Therefore, for any $\theta>0,~\epsilon>0$ and $j$ there exists a $B_{j,j^\prime}$ such that for all $b\geq B_{j,j^\prime}$
$$\Pr[\big|\big\{\xi_{j^\prime b}(\bZ)-\xi_{j b}(\bZ)\big\}-\{\tau_{j^\prime b}(\bZ)-\tau_{jb}(\bZ)\} - \{\tilde{\xi}_b(j,j^\prime)-\tilde{\tau}_b(j,j^\prime)\}\big|<\theta\big | \bZ\sim F_j]>1-\epsilon.$$

We assume $\tilde{\xi}_b(j,j^\prime)>\tilde{\tau}_b(j,j^\prime)$ for all $b\geq B_1$ and $1 \leq j\neq j^\prime \leq J.$ Let $\theta_0 = \lim\inf_b \big (\tilde{\xi}_b(j,j^\prime)-\tilde{\tau}_b(j,j^\prime) \big )$. By assumption $(A9)$, $\theta_0 > 0$. Hence, for any  $0<\theta < \theta_0$ and $\epsilon > 0$, there exists a $b^\prime(\theta_0,\theta,\epsilon)$ such that for all $b\geq b^\prime(\theta_0,\theta,\epsilon)$
$$\Pr[\big\{\xi_{j^\prime b}(\bZ)-\xi_{j b}(\bZ)\big\}-\{\tau_{j^\prime b}(\bZ)-\tau_{jb}(\bZ)\} \leq 0\big | \bZ\sim F_j]<\epsilon.$$
From equation \eqref{thm3.5}, we now obtain
\begin{align*}
&~\Pr[\tau_{j^\prime b}(\bZ)-\tau_{jb}(\bZ)>0~\forall j \neq j^\prime|\bZ\sim \mathbf{F}_j]\\
& \leq {\rm E}\bigg \{\prod\limits_{1\leq j\neq j^\prime\leq J} \Pr[\xi_{j^\prime b}(\bZ)-\xi_{j b}(\bZ)>0|\bZ\sim \mathbf{F}_j,\rchi_j]\bigg \}+\sum\limits_{1\leq j\neq j^\prime\leq J}C_J~\epsilon\\
& = {\rm E}\bigg \{\prod\limits_{1\leq j\neq j^\prime\leq J} \Pr[\xi_{j^\prime b}(\bZ)-\xi_{j b}(\bZ)>0|\bZ\sim \mathbf{F}_j,\rchi_j]\bigg \} + C^{\prime}_J~\epsilon\\
& = {\rm E}\big \{ \Pr[\xi_{j^\prime b}(\bZ)-\xi_{j b}(\bZ)>0~\forall j^\prime\neq j, 1\leq j^\prime \leq J|\bZ\sim \mathbf{F}_j,\rchi_j]\big \} + C^{\prime}_J~\epsilon\\[5pt]
& = \Pr[\xi_{j^\prime b}(\bZ)-\xi_{j b}(\bZ)>0~\forall j^\prime\neq j, 1\leq j^\prime \leq J|\bZ\sim \mathbf{F}_j] + C^{\prime}_J~\epsilon~\text{ for all } b \geq b^\prime(\theta_0,\theta,\epsilon).
\end{align*}
Therefore,
\begin{align*}
&\sum_{j=1}^J \pi_j \Pr[\tau_{j^\prime b}(\bZ)-\tau_{jb}(\bZ)>0~\forall j^\prime\neq j, 1\leq j^\prime \leq J|\bZ\sim \mathbf{F}_j]\\
&~\leq \sum_{j=1}^J \pi_j \Pr[\xi_{j^\prime b}(\bZ)-\xi_{j b}(\bZ)>0~\forall j^\prime\neq j, 1\leq j^\prime \leq J|\bZ\sim \mathbf{F}_j]+C^{\prime}_J~\epsilon\\
\implies&\Pr[\delta_{\rm NN-bgMADD}(\bZ)=Y]\leq \Pr[\delta_{\rm bgSAVG}(\bZ)=Y]+C^{\prime}_J~\epsilon.
\end{align*}
This now implies that $\Delta_{\rm bgSAVG}-C^{\prime}_J~\epsilon \leq \Delta_{\rm NN-bgMADD}$ for all $b\geq b^\prime(\theta_0,\theta,\epsilon)$. Since $\epsilon>0$ is arbitrarily, we conclude that 
$$\Delta_{\rm bgSAVG} \leq \Delta_{\rm NN-bgMADD}\text{ for all }b\geq b^\prime(\theta_0,\theta,\epsilon).$$ 
Following a similar line of arguments, one can prove that there exist $B_1$ and $B_2$ such that if $\tilde{\xi}_b(j,j^\prime)<\tilde{\tau}_b(j,j^\prime)$ for all $b\geq B_1$ and $1 \leq j\neq j^\prime \leq J$, then $\Delta_{\rm bgSAVG} \geq \Delta_{\rm NN-bgMADD}$ for all $b\geq B_2$. This completes the proof.
\QEDB

\begin{lemma} \label{xi_better_than_psi}
We now discuss some sufficient conditions for $\tilde{\xi}^b_{\phi,\gamma}(j,j^\prime)\geq (<) ~\tilde{\tau}^b_{\phi,\gamma}(j,j^\prime)$ for $1 \leq j\neq j^\prime \leq J$.\\ 
Let us consider a two ($J=2$) class problem. If
\begin{enumerate}[i.]
	\item $\tilde{h}_b(1,2)>\tilde{h}_b(1,1)>\tilde{h}_b(2,2)$ and $n_1> n_2+1$,
	\item $\tilde{h}_b(1,2)>\tilde{h}_b(2,2)>\tilde{h}_b(1,1)$ and $n_1< n_2 -1$, 
	\item $\tilde{h}_b(1,1)>\tilde{h}_b(1,2)\geq\frac{3}{4}\tilde{h}_b(1,1)+ \frac{1}{4}\tilde{h}_b(2,2)>\tilde{h}_b(2,2)$ and 
	\item[] $n_1> 1+\frac{n-1}{2} \Big \{ \displaystyle \frac{\tilde{h}_b(1,1)-\tilde{h}_b(2,2)}{2\tilde{h}_b(1,2)-\tilde{h}_b(1,1)-\tilde{h}_b(2,2)} \Big \}$, or
	\item $\tilde{h}_b(2,2)>\tilde{h}_b(1,2)\geq\frac{1}{4}\tilde{h}_b(1,1)+ \frac{3}{4}\tilde{h}_b(2,2)>\tilde{h}_b(1,1)$ and
	\item[] $n_1< (n-1) \Big \{ \displaystyle 1 - \frac{1}{2}\frac{\tilde{h}_b(2,2)-\tilde{h}_b(1,1)}{2\tilde{h}_b(1,2)-\tilde{h}_b(1,1)-\tilde{h}_b(2,2)} \Big \},$ 
\end{enumerate}
then $\tilde{\xi}_b(1,2)> \max\{\tilde{\tau}_b(1,2),\tilde{\tau}_b(2,1)\}$.
\end{lemma}

\noindent
{\bf Proof of Lemma \ref{xi_better_than_psi}} Please check Section 1 of the Supplementary for a proof.
\QEDB
\vspace{0.1in}

\noindent
{\bf Remark A \label{A8_const}}
Assumption $(A8)$ holds in various scenarios. In particular, if the component variables of the underlying distributions are i.i.d., then the constants $\tilde{\xi}_b$ and $\tilde{\tau}_b$ are free of $b$. To realize this, assume $\bX_1,\bX_2\stackrel{i.i.d}{\sim}{\bf F}_1,\ \bX_2,\bX_4\stackrel{i.i.d}{\sim}{\bf F}_2$. If $d_i=d_1$, and $\bX_{1i}\stackrel{i.i.d}{\sim}{\bf F}_{1,i}$, $\bX_{3i}\stackrel{i.i.d}{\sim}{\bf F}_{2,i} \text{ for all } 1 \leq i \leq b$, then we have
\begin{align*}
\tilde{h}_b(1,2)&=\phi \bigg (\frac{1}{b}\sum\limits_{i=1}^b {\rm E}[\gamma(\frac{1}{d_i}\|\bX_{1i}-\bX_{3i}\|^2)]\bigg )\\
&=\phi \bigg (\frac{1}{b}\sum\limits_{i=1}^b {\rm E}[\gamma(\frac{1}{d_1}\|\bX_{11}-\bX_{31}\|^2)]\bigg )\\
&=\phi \bigg ({\rm E}[\gamma(\frac{1}{d_1}\|\bX_{11}-\bX_{31}\|^2)]\bigg ),
\end{align*}
which implies that $\tilde{h}_b(1,2)$ is free of $b$. Similarly, we can show that
$\tilde{h}_b(1,1) = \phi \big ({\rm E}[\gamma(\frac{1}{d_1}\|\bX_{11}-\bX_{21}\|^2)]\big )$ and $\tilde{h}_b(2,2) = \phi \big ({\rm E}[\gamma(\frac{1}{d_1}\|\bX_{31}-\bX_{41}\|^2)]\big )$ are also free of $b$. Consequently,\\ $\liminf_b{\tilde{\xi}_b(1,2)} (= \tilde{\xi}_1(1,2)$, say) and $\liminf_b{\tilde{\tau}_b(1,2)} (=\tilde{\tau}_1(1,2)$, say) remain constant for varying $b$. Clearly, under such circumstances, a sufficient condition for assumption $(A8)$ is
$$|\tilde{\xi}_1(1,2)- \tilde{\tau}_1(1,2)|>0.$$

\noindent
It is also straightforward to observe that if ${\rm E}[\gamma(d_i^{-1}\|{\bf U}_{i}-{\bf V}_{i}\|^2)]={\rm E}[\gamma(d_{i^\prime}^{-1}\|{\bf U}_{i^\prime}-{\bf V}_{i^\prime}\|^2)]$ for all $1\leq i,i^\prime\leq b$, with ${\bf U} \sim \mathbf{F}_j$ and ${\bf V} \sim \mathbf{F}_{j^\prime}$, then both $\tilde{\xi}_b(j,j^\prime)$ and $\tilde{\tau}_b(j,j^\prime)$ are also free of $b$. 
\QEDB


\begin{lemma} \label{mixinglemma}
Suppose $\bU=\{\bU_1,\bU_2,\ldots\}$ and $\bV=\{\bV_1,\bV_2,\ldots\}$ with $\bU_i$ and $\bV_i$ denoting the respective sub-vectors for $i \in \mathbb{N}$. If $\bU$ and $\bV$ are $\rho$-mixing sequences, then the sequence $\bW=(W_1,W_2,\ldots)^\top$, where $W_i=\gamma({d_i}^{-1}\|\bU_i-\bV_i\|^2)$, is $\rho$-mixing and $\mathop{\sum\sum}_{1\leq i< i^\prime\leq b} {\rm Corr}\left(W_i,W_{i^\prime}\right)\\ = o(b^2)$.
\end{lemma}

\noindent
{\bf Proof of Lemma \ref{mixinglemma}}
For a random sequence $\bX=(X_1,X_2,\ldots)^{\top}$ we have
$$\rho_{\bX}(d)=\sup_{k\geq1} \rho \big (\sigma(X_1,\ldots,X_k),\sigma(X_{k+d},\ldots) \big ),$$
where
$\sigma(X_i, i \in I)\text{ denotes the }\sigma\text{-field generated by }\{X_i, i \in I\},\text{ and }
\rho(\mathcal{A},\mathcal{B}) \mbox{ is defined as }$ $\sup_{X\in \mathscr{L}^2(\mathcal{A}),Y\in \mathscr{L}^2(\mathcal{B})}\big|{\rm E}[XY]-{\rm E}[X]{\rm E}[Y]\big|.$
Here, $\mathscr{L}^2(\mathcal{A})$ is the space of square integrable random variables on $\mathcal{A}$. The sequence $\bX$ is said to be $\rho$-mixing if $\rho_{\bX}(d) \to 0$ as $d\to\infty$ \citep[see, e.g.,][]{Bradley2007}.

Define $Z_i=h(U_i,V_i)$ for $i \in \mathbb{N}$, where $h:\mathbb{R}^2\to\mathbb{R}$ is a continuous function. Note that $\sigma(Z_{a_1},\ldots,Z_{a_2})\subseteq\sigma(U_{a_1},\ldots,U_{a_2})\vee\sigma(V_{a_1},\ldots,V_{a_2})$. \cite{Bradley2007} showed that
\begin{align}\label{mixrel1}
\rho_{\bZ}(d) & = \sup_{k\geq 1}\rho\big(\sigma(Z_1,\ldots,Z_k),\sigma(Z_{k+d},\ldots)\big) \\
& \leq \sup_{k\geq 1}\rho\big(\sigma(U_1,\ldots,U_k)\vee\sigma(V_1,\ldots,V_k),\sigma(U_{k+d},\ldots)\vee\sigma(V_{k+d},\ldots)\big) \nonumber\\ 
&(\text{see Theorem 3.15-Remark (I), p.82 of \citealp{Bradley2007}}) \nonumber\\
&=\sup_{k\geq 1}\max~\{\rho\big(\sigma(U_1,\ldots,U_k),\sigma(U_{k+d},\ldots)\big),~\rho\big(\sigma(V_1,\ldots,V_k),\sigma(V_{k+d},\ldots)\big)\} \nonumber \\
& (\text{see Theorem 6.6-(II) and Note 3, pp.199-200 of \citealp{Bradley2007}}) \nonumber\\
& = \max~\{\sup_{k\geq 1}\rho\big(\sigma(U_1,\ldots,U_k),\sigma(U_{k+d},\ldots)\big),~\sup_{k\geq 1}\rho\big(\sigma(V_1,\ldots,V_k),\sigma(V_{k+d},\ldots)\big)\}\nonumber\\
& = \max~\{\rho_{\bU}(d),\rho_{\bV}(d)\}.
\end{align}
Therefore, $\rho_{\bZ}(d) \to 0$ if both $\rho_{\bU}(d) \to 0$ and $\rho_{\bV}(d) \to 0$ as $d \to \infty$. 

\vspace{0.5cm}
Let us consider the sequence $\bW$ with $W_1=g_1(Z_1,\ldots,Z_{d_1})$, $W_2=g_2(Z_{d_1+1},\ldots,Z_{d_1+d_2})$ and so on, where $g_i:\mathbb{R}^{d_i}\to\mathbb{R}$ for $i \in \mathbb{N}$ are continuous functions. For simplicity, let us assume that $d_i=d_0$ for all $1 \leq i \leq b$. Now, we have
\begin{align*}
~\sigma(W_{a_1},\ldots,W_{a_2})
=&~\sigma(g_{a_1}(Z_{(a_1-1)d_0+1},\ldots,Z_{a_1d_0}),\ldots,g_{a_2}(Z_{(a_2-1)d_0+1},\ldots,Z_{a_2d_0}))\\
\subseteq&~\sigma(Z_{(a_1-1)d_0+1},\ldots,Z_{a_1d_0},\ldots,Z_{(a_2-1)d_0+1},\ldots,Z_{(a_2-1)d_0}).
\end{align*}
This further implies that
\begin{align}\label{mixrel2}
\rho_{\bW}(d) &=\sup_{k\geq 1}\rho\big(\sigma(W_1,\ldots,W_k),\sigma(W_{k+d},\ldots)\big) \\
&\leq \sup_{k\geq 1}\rho\big(\sigma(Z_1,\ldots,Z_{d_0},\ldots,Z_{(k-1)d_0+1},\ldots,Z_{kd_0}),\sigma(Z_{(k+d-1)d_0+1},\ldots,Z_{(k+d)d_0},\ldots)\big) \nonumber\\
& (\text{see Theorem 3.15-Remark (I), p.82 of \citealp{Bradley2007}}) \nonumber\\
&\leq \sup_{k\geq 1}\rho\big(\sigma(Z_1,\ldots,Z_{d_0},\ldots,Z_{(k-1)d_0+1},\ldots,Z_{kd_0}),\sigma(Z_{kd_0+d},Z_{kd_0+d+1}\ldots)\big)\nonumber\\
&= \sup_{k\geq 1}\rho\big(\sigma(Z_1,\ldots,Z_k),\sigma(Z_{k+d},\ldots)\big)\nonumber\\
&= \rho_{\bZ}(d).
\end{align}
Proof for the case when $d_i$s are {\it unequal, but bounded} follows by using a similar line of arguments. From equations (\ref{mixrel1}) and (\ref{mixrel2}), it follows that $\bW$ is a $\rho$-mixing sequence if both the original sequences $\bU$ and $\bV$ are $\rho$-mixing.
Consider the maps $h(u,v)=(u-v)^2$, $g(u_1,\ldots,u_k)=(u_1+\cdots+u_k)/k$, and $\gamma$ as described in Lemma \ref{thirdlemma}.
Hence, if $\bU$ and $\bV$ are $\rho$-mixing, then the sequence 
$\bW=\{W_i=\gamma({d_i}^{-1}\|\bU_i-\bV_i\|^2),i\geq 1\}$ is also $\rho$-mixing.

Now, by Theorem 4.5(b) of \cite{Bradley2007}, we have 
$${\rm Corr}(W_i,W_{i^\prime}) \leq \rho(\sigma(W_i),\sigma(W_{i^\prime})) \leq \rho\big(\sigma(W_1,\ldots,W_i),\sigma(W_{i^\prime},\ldots)\big) \leq \rho_{\bW}(i^\prime-i).$$ 
Therefore,
$$0\leq  b^{-2} \mathop{\sum\sum}_{1\leq i<i^\prime\leq b} {\rm Corr}(W_i,W_{i^\prime}) \leq b^{-2} \mathop{\sum\sum}_{1\leq i<i^\prime\leq b} \rho_{\bW}(i^\prime-i) \leq b^{-2} \sum_{l=1}^b (b-l) \rho_{\bW}(l) \leq b^{-1} \sum_{l=1}^b \rho_{\bW}(l).$$
Since, $\rho_{\bW}(b) \to 0$ as $b \to \infty$, it follows from Ces{\`a}ro summability that $$\mathop{\sum\sum}_{1\leq i< i^\prime\leq b} {\rm Corr}\left(W_i,W_{i^\prime}\right) = o(b^2).$$ \QEDB



\newpage
\section{Notations} \label{AppendixB}

\begin{table}[h]
	\centering
	\caption{List of standard notations}
	\vspace{0.1in}
	\begin{tabular}{|c|c|c|c|}
		\hline
		\multicolumn{1}{|c|}{\it Symbol} & \multicolumn{1}{|c|}{\it Denotes} \\
		\hline
		\multicolumn{1}{|c|}{$J$} & number of classes \\
		\hline
		\multicolumn{1}{|c|}{$n_j$} & training sample size of $j$-th class \\
		\hline
		\multicolumn{1}{|c|}{$n$} & total training sample size  \\
		\hline
		\multicolumn{1}{|c|}{$d$} & data dimension \\
		\hline
		\multicolumn{1}{|c|}{$\rchi$} & random sample \\
		\hline
		\multicolumn{1}{|c|}{$\bmu$} & location parameter (vector) \\
		\hline
		\multicolumn{1}{|c|}{$\Sigma$} & scale parameter (matrix) \\
		\hline
		\multicolumn{1}{|c|}{$\rho$} & population correlation coefficient \\
		\hline
		\multicolumn{1}{|c|}{$r$} & sample correlation coefficient \\
		\hline
		\multicolumn{1}{|c|}{$X$} & random variable \\
		\hline
		\multicolumn{1}{|c|}{${\bf X}$} & random vector \\
		\hline
		\multicolumn{1}{|c|}{$F$} & distribution function of a random variable $X$ \\
		\hline
		\multicolumn{1}{|c|}{${\bf F}$} & distribution function of a random vector ${\bf X}$ \\
		\hline
		\multicolumn{1}{|c|}{${\rm Corr}(X,Y)$} & correlation between $X$ and $Y$ \\
		\hline
		\multicolumn{1}{|c|}{$\delta$} & a generic classifier \\
		\hline
		\multicolumn{1}{|c|}{$\Delta$} & misclassification probability (rate) of the classifier $\delta$ \\
		\hline
	\end{tabular}%
	\label{symtab1}%
\end{table}%

\vspace{0.1in}

\begin{table}[h]
	\centering
	\caption{Notations specific to this paper}
	\vspace{0.1in}
	\begin{tabular}{|c|c|c|c|c|}
		\hline
		\multicolumn{1}{|c|}{\it Symbol} & \multicolumn{1}{|c|}{\it Denotes} & \multicolumn{1}{|c|}{\it Remark} \\
		\hline
		\multicolumn{1}{|c|}{$b$} & number of blocks  & \\
		\hline
		\multicolumn{1}{|c|}{$h({\bf U},{\bf V})$} &  generalized distance between ${\bf U}$ and ${\bf V}$ & \\
		\hline
		\multicolumn{1}{|c|}{$\xi({\bf U},{\bf V})$} & measure of dissimilarity between ${\bf U}$ and ${\bf V}$ & average distance classifier \\
		\hline
		\multicolumn{1}{|c|}{$\tilde{\xi}(j,j^\prime)$} & measure of separability between class $j$ and $j^\prime$ & average distance classifier \rule{0pt}{2.6ex} \\
		\hline
		\multicolumn{1}{|c|}{$\tau({\bf U},{\bf V})$} & measure of dissimilarity between ${\bf U}$ and ${\bf V}$ & nearest neighbor classifier \\
		\hline
		\multicolumn{1}{|c|}{$\tilde{\tau}(j,j^\prime)$} & measure of separability between class $j$ and $j^\prime$ & nearest neighbor classifier \\
		\hline
	\end{tabular}%
	\label{symtab2}%
\end{table}
\vspace{0.05in}

\newpage
\bibliography{citation}


\end{document}